\documentclass[3p,preprint,number]{elsarticle}
\DeclareGraphicsExtensions{.pdf,.gif,.jpg,.pgf}

\usepackage[english]{babel}
\usepackage{ragged2e}
\usepackage{blindtext}
\usepackage{subcaption}
\usepackage{mathtools}
\usepackage{lipsum}
\usepackage{array}
\usepackage{colortbl}
\usepackage{chemfig}
\usepackage{tikz}
\usepackage{pgfplots}
\usepackage{amsmath}
\usepackage{physics}
\pgfplotsset{compat=1.17}
\usepackage{amssymb}
\usepackage{subcaption}
\usepackage{siunitx}
\usepackage{enumitem}
\usepackage{xcolor}
\usepackage{multirow}
\usepackage{adjustbox}
\usepackage[latin1]{inputenc}
\usepackage{csquotes}
\usetikzlibrary{patterns}
\usepackage[english]{babel}

\makeatletter
\newcommand*{\rom}[1]{\expandafter\@slowromancap\romannumeral #1@}
\makeatother
\usepackage{tikz}
\usetikzlibrary{matrix}
\graphicspath{ {c:/Users/User/Documents/PhD/Year 1/LaTeX/Figures/} }
\usepackage{blindtext}
\usepackage{titlesec}
\titleformat{\chapter}[display]   
{\normalfont\huge\bfseries}{\chaptertitlename\ \thechapter}{16pt}{\Huge}   
\titlespacing*{\chapter}{0pt}{-10pt}{5pt}
\usepackage[longend,ruled]{algorithm2e}
\usepackage{algorithmicx} 
\usepackage{layouts}

\usepackage{changes} 
\definechangesauthor[name=James Mckenna, color=blue]{JM}

\begin{document}
	\bibliographystyle{elsarticle-num.bst}
	{\setlength{\parindent}{0cm} 
		\begin{frontmatter}
			\title{A Local Multi-Layer Approach to Modelling Interactions between Shallow Water Flows and Obstructions}
			\author{James Mckenna\corref{cor1}}
			\ead{J.Mckenna4@newcastle.ac.uk}
			\author{Vassilis Glenis}
			\ead{Vassilis.Glenis@newcastle.ac.uk}
			\author{Chris Kilsby}
			\ead{Chris.Kilsby@newcastle.ac.uk}
			
			\cortext[cor1]{Corresponding author}
			\address{School of Engineering, Newcastle University, Newcastle upon Tyne, United Kingdom}
			\date{Submitted April 2023}
			\begin{abstract}				
				The capability to accurately predict flood flows via numerical simulations is a key component of contemporary flood risk management practice. However, modern flood models lack the capacity to accurately model flow interactions with linear features, or hydraulic structures like bridges and gates, which act as partial barriers to flow. Presented within this paper is a new Riemann solver which represents a novel approach to modelling fluid-structure interactions within two-dimensional hydrodynamic models. The solution procedure models obstacles as existing at the interface between neighbouring cells and uses a combination of internal boundary conditions, different forms of the conservation laws and vertical discretisation of the neighbouring cells to resolve numerical fluxes across a partially obstructed interface. The predictive capacity of the solver has been validated through comparisons with experimental data collected from experiments conducted in a state-of-the-art hydraulic flume. Since the solution procedure is local, only applying to the cells within the immediate vicinity of a structure, the method is designed to be compatible with existing two-dimensional hydrodynamic models which use a finite volume scheme to solve the shallow water equations. 
			\end{abstract}
			\begin{keyword}
				Flood modelling; bridges; free-surface flow; Riemann solver; finite-volume; model validation.
			\end{keyword}
		\end{frontmatter}
		\section{Introduction}
		The ominous threat of anthropogenic climate change is driving the requirement for more effective flood risk management in order to better manage what is already a challenging and costly hazard; models estimate that forecasted average annual flood losses for the United States will increase from US\$32 billion to more than US\$40 billion by 2050 \cite{RN302}, with similar predictions of increasing flood risk being made on a global scale \cite{RN303}. Hydrodynamic models play a vital role in contemporary flood risk management by providing evidence, via numerical predictions, upon which the quantification of flood risk and consequential future investment is based. It is therefore vital for effective flood risk management that hydrodynamic models produce accurate predictions. 
		
		Within catchments, channel structures, such as bridges, weirs and gates, can act as obstacles to flow, significantly influencing the local flow characteristics \cite{RN90}. However, within modern hydrodynamic modelling practice, methods for modelling such features are relatively under-developed, with industry standard models using coarse approximations, empirically based methods or even omitting such features entirely \cite{RN265,RN272,RN273}. Within academic literature there have been a number of contributions towards bridging this gap in modelling capacity, such as \cite{RN102, RN107, RN376, RN101, RN232}, however, none of the published works present an accurate method for the generalised treatment of partial barriers to flow within two-dimensional hydrodynamic models. 
		
		Within Mckenna et al. \cite{RN100}, the authors of this paper presented a new Riemann solver capable of resolving numerical fluxes across a partially obstructed interface. The proposed solution procedure represents structures as existing at the interface between neighbouring cells and uses a combination of internal boundary conditions and a different form of the conservation laws in the adjacent cells, to resolve numerical fluxes across the partially obstructed interface. Experimental validation, via experiments conducted in a state of the art research flume, demonstrated the accuracy of the solver for a range of flow conditions and barrier configurations. 
		
		Despite the successful validation of the solver, there is opportunity for enhancement of the method via more accurate discretisation of the horizontal velocity in the vertical plane. As such, this paper aims to use the basic conceptual idea underpinning the Riemann solver developed in \cite{RN100}, which is the decomposition of the Riemann problem in the vertical plane, to develop a new, more sophisticated and accurate method for representing structures within two-dimensional hydrodynamic models. As for the development of the previous solver, compatibility of the method with existing flood models utilising two dimensional finite volume schemes to solve the shallow water equations was a key consideration throughout the development of the solver.
		\section{Mathematical Model}
		The proposed solution method divides the computational domain into structure cells, intermediate cells and normal cells with corresponding normal interfaces (NI), intermediate interfaces (II) and structure interfaces (SI) as shown in Figure \ref{fig: Cells}.
		\begin{figure}[hbt!]
			\centering
			\begin{tikzpicture}
				\draw[thick, ->] (-0.75,0) -- (-0.75,2.5) node[above] {$h$};
				
				\draw[thick] (0,0) -- (12,0); 
				
				\fill[fill=blue!25!white, draw=black] (0,0) rectangle (2,2); 
				\draw[thick] (0,-0.5) -- (0,2.5) node[above] {$a$}; 
				\draw[dashed] (2,-0.5) -- (2,2.5) node[above] {NI}; 
				\fill[fill=blue!40!white, draw=black] (2,0) rectangle (4,2); 
				\draw[dashed] (4,-0.5) -- (4,2.5) node[above] {II}; 
				\draw[dashed] (6,-0.5) -- (6,2.5) node[above] {SI}; 
				
				\fill[fill=blue!55!white, draw=black] (4,0) rectangle (6,2); 
				\fill[fill=blue!55!white, draw=black] (6,0) rectangle (8,2); 
				\fill[fill=black!25!white, draw=black] (5.9,0.5) rectangle (6.1,1.5); 
				
				\draw[dashed] (8,-0.5) -- (8,2.5) node[above] {II}; 
				\fill[fill=blue!40!white, draw=black] (8,0) rectangle (10,2); 
				\draw[dashed] (10,-0.5) -- (10,2.5) node[above] {NI}; 
				\fill[fill=blue!25!white, draw=black] (10,0) rectangle (12,2); 
				\draw[thick] (12,-0.5) -- (12,2.5) node[above] {$b$}; 
				
				\draw[thick, dashed] (3.25, 0.5) -- (8.75, 0.5) node[right] {\large$z_{\frac{1}{2}}$};
				\draw[thick, dashed] (3.25, 1.5) -- (8.75, 1.5) node[right] {\large$z_{\frac{3}{2}}$};
				\draw [decorate, decoration={brace,amplitude=6pt,raise=0pt}, thick] (8,-0.5) -- (4,-0.5); 
				\node[below] at (6,-0.65) {\large Structure Cells};
				\fill[fill=black, draw=black] (5,0) circle (0.1cm);
				\node[below] at (5,-0.1) {$x_{i}$};
				\fill[fill=black, draw=black] (7,0) circle (0.1cm);
				\node[below] at (7,-0.1) {$x_{i+1}$};
				\draw [decorate, decoration={brace,amplitude=6pt,raise=0pt}, thick] (4,-0.5) -- (2,-0.5); 
				\node[below] at (3,-0.65) {\large Intermediate};
				\node[below] at (3,-1.15) {\large Cell};
				\fill[fill=black, draw=black] (3,0) circle (0.1cm);
				\node[below] at (3,-0.1) {$x_{i-1}$};
				\draw [decorate, decoration={brace,amplitude=6pt,raise=0pt}, thick] (2,-0.5) -- (0,-0.5); 
				\node[below] at (1,-0.65) {\large Normal};
				\node[below] at (1,-1.15) {\large Cell};
				\fill[fill=black, draw=black] (1,0) circle (0.1cm);
				\node[below] at (1,-0.1) {$x_{i-2}$};
				\draw [decorate, decoration={brace,amplitude=6pt,raise=0pt}, thick] (10,-0.5) -- (8,-0.5); 
				\node[below] at (9,-0.65) {\large Intermediate};
				\node[below] at (9,-1.15) {\large Cell};
				\fill[fill=black, draw=black] (9,0) circle (0.1cm);
				\node[below] at (9,-0.1) {$x_{i+2}$};
				\draw [decorate, decoration={brace,amplitude=6pt,raise=0pt}, thick] (12,-0.5) -- (10,-0.5); 
				\node[below] at (11,-0.65) {\large Normal};
				\node[below] at (11,-1.15) {\large Cell};
				\fill[fill=black, draw=black] (11,0) circle (0.1cm);
				\node[below] at (11,-0.1) {$x_{i+3}$};
			\end{tikzpicture}
			\caption{A simple computational domain $[a,b]$ illustrating the designation of structure, intermediate and normal cells with their corresponding interfaces. $z_{1/2}$ and $z_{3/2}$ represent the height above the bed of the base and cover of the idealised structure represented at the structure interface.}
			\label{fig: Cells}
		\end{figure}
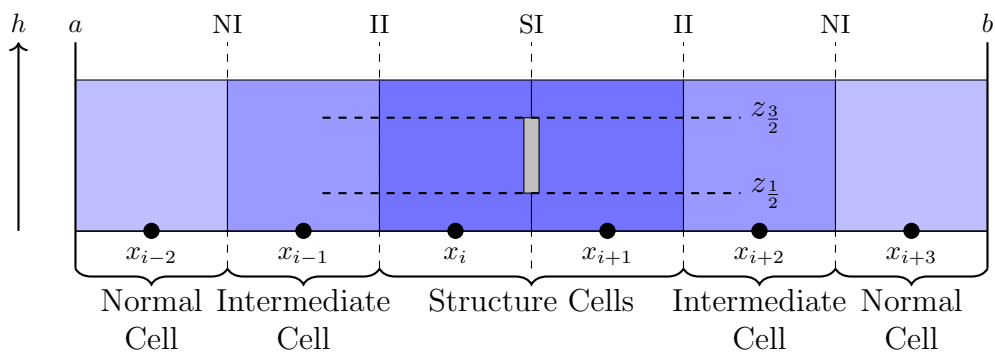 \\
		At a structure interface, the adjacent structure cells are vertically discretised into sub-cells with a maximum depth capacity corresponding to the dimensions of the idealised structure represented at the interface as shown in Figure \ref{fig: Structure Cells}. For example, the sub-cells $\textbf{U}_{i,1}$ and $\textbf{U}_{i+1,1}$ in Figure \ref{fig: Structure Cells} have a maximum depth capacity of $h_1 = z_{\frac{3}{2}} - z_b$, which represents the difference in elevation between the base of the structure and the bed.
		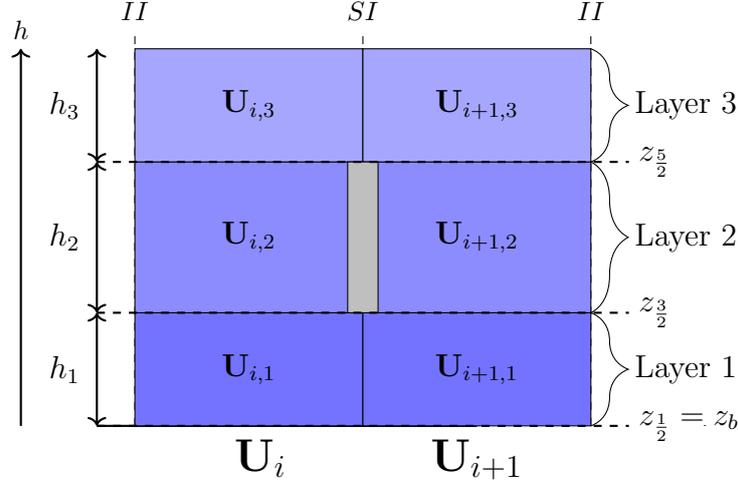
\begin{figure}[hbt!]
			\centering
			\begin{tikzpicture}
				\draw[thick, ->] (2.5,0) -- (2.5,5) node[above] {$h$};
				\draw[thick] (11.5,0) -- (11.5,0.01);
				\draw[thick] (3.5,0) -- (8.5,0); 
				\draw[dashed] (4,0) -- (4,5.25) node[above] {$II$};
				\draw[dashed] (7,0) -- (7,5.25) node[above] {$SI$};
				\draw[dashed] (10,0) -- (10,5.25) node[above] {$II$};
				
				\fill[fill=blue!55!white, draw=black] (4,0) rectangle (7,1.5); 
				\fill[fill=blue!45!white, draw=black] (4,1.5) rectangle (7,3.5); 
				\fill[fill=blue!35!white, draw=black] (4,3.5) rectangle (7,5); 
				\fill[fill=blue!55!white, draw=black] (7,0) rectangle (10,1.5); 
				\fill[fill=blue!45!white, draw=black] (7,1.5) rectangle (10,3.5); 
				\fill[fill=blue!35!white, draw=black] (7,3.5) rectangle (10,5); 
				\fill[fill=black!25!white, draw=black] (6.8,1.5) rectangle (7.2,3.5); 
				
				\draw[thick,<->] (3.5,0) -- (3.5,1.5);
				\node[left] at (3.4, 0.75) {\large$h_1$};
				\draw[thick,<->] (3.5,1.5) -- (3.5,3.5);
				\node[left] at (3.4, 2.5) {\large$h_2$};
				\draw[thick,<->] (3.5,3.5) -- (3.5,5);
				\node[left] at (3.4, 4.25) {\large$h_3$};
				
				\draw[thick, dashed] (3.5, 0) -- (10.5, 0) node[right] {\large$z_{\frac{1}{2}} = z_b$};
				\draw[thick, dashed] (3.5, 1.5) -- (10.5, 1.5) node[right] {\large$z_{\frac{3}{2}}$};
				\draw[thick, dashed] (3.5, 3.5) -- (10.5, 3.5) node[right] {\large$z_{\frac{5}{2}}$};
				\draw [decorate,decoration={brace,amplitude=14pt},xshift=0pt,yshift=0pt] (10,1.5) -- (10,0) node [black,midway,xshift=1.25cm] {\large Layer 1};
				\draw [decorate,decoration={brace,amplitude=14pt},xshift=0pt,yshift=0pt] (10,3.5) -- (10,1.5) node [black,midway,xshift=1.25cm] {\large Layer 2};
				\draw [decorate,decoration={brace,amplitude=14pt},xshift=0pt,yshift=0pt] (10,5) -- (10,3.5) node [black,midway,xshift=1.25cm] {\large Layer 3};
				
				\node[below] at (5.65,-0.05) {\LARGE $\textbf{U}_{i}$};
				\node[below] at (8.5,-0.05) {\LARGE $\textbf{U}_{i+1}$};
				
				\node[] at (5.5,0.75) {\large $\textbf{U}_{i,1}$};
				\node[] at (8.5,0.75) {\large $\textbf{U}_{i+1,1}$};
				\node[] at (5.5,2.5) {\large $\textbf{U}_{i,2}$};
				\node[] at (8.5,2.5) {\large $\textbf{U}_{i+1,2}$};
				\node[] at (5.5,4.25) {\large $\textbf{U}_{i,3}$};
				\node[] at (8.5,4.25) {\large $\textbf{U}_{i+1,3}$};

			\end{tikzpicture}
			\caption{Division of structure cells into sub-cells corresponding to the base and cover of the idealised structure modelled at the structure interface. The maximum depth capacity of flow in the layer one is $z_{3/2}-z_{1/2}$ and the maximum depth capacity of flow in the second layer is equal to $z_{5/2}-z_{3/2}$. The uppermost layer has no maximum depth capacity.}
			\label{fig: Structure Cells}
		\end{figure}
	
		For normal interfaces and the corresponding adjacent normal or intermediate cells, a one-dimensional (1D) FV scheme is used to solve the 1D Shallow Water Equations (1D-SWE) given as:
		\begin{equation}
			\partial_t\textbf{U} + \partial_x\textbf{F}(\textbf{U})=\textbf{S}(\textbf{U})
		\end{equation}
		Where $\textbf{U}$ is the vector of conserved variables, $\textbf{F}(\textbf{U})$ is the vector of fluxes and $\textbf{S}(\textbf{U})$ is a vector of sources comprising of $\textbf{S}_0$, the bed slope source term and $\textbf{S}_f$, the bed friction source term. These terms are given as follows:
		\begin{gather}
			\textbf{U} = \begin{bmatrix}
				h \\[6pt]
				hu \\
			\end{bmatrix} \; \textrm{ , } \;
			\textbf{F} = \begin{bmatrix}
				hu \\[6pt]
				hu^2 + \frac{1}{2}gh^2 \\
			\end{bmatrix} \; \textrm{ , } \;
			\textbf{S}_0 = \begin{bmatrix}
				0 \\[6pt]
				-gh\frac{\partial z}{\partial x} \\
			\end{bmatrix} \; \textrm{ , } \;
			\textbf{S}_f = \begin{bmatrix}
				0 \\[6pt]
				- \tau_f \\
			\end{bmatrix}
		\end{gather}
		Whereby $h$ denotes the depth of flow, $u$ denotes the velocity component in the $x$ direction, $g$ is the acceleration due to gravity, $z$ is the elevation of the bed and $\tau_f$ is the shear stress due to bed friction in accordance with Manning's equation:
		\begin{equation}\label{eq: Manning's Friction}
			\tau_f = C_fu|u| = \frac{gn^2}{\sqrt[3]{h}}u|u|
		\end{equation}
		Where $n$ is Manning's roughness coefficient.
		
		For the structure and intermediate interfaces and corresponding adjacent structure and intermediate cells, a 1D FV scheme is used to solve a multi-layer 1D shallow water system \cite{RN39}:
		\begin{equation}\label{eq: 2-layer SWE 1}
			\partial_t\textbf{U}_k + \partial_x\textbf{F}_k(\textbf{U}_k)=\textbf{S}_k(\textbf{U}_k)
		\end{equation}
		Where $\textbf{U}_k$ is the vector of conserved variables for the layer $k$, $\textbf{F}(\textbf{U}_k)$ is the vector of fluxes for layer $k$ and $\textbf{S}_k(\textbf{U}_k)$ is a vector of sources for layer $k$ comprising of $\textbf{S}_{k,0}$, the topographic source terms for layer $k$ and $\textbf{S}_{k,f}$, the friction source terms for layer $k$. These terms are given as follows:
		\begin{gather}
			\textbf{U}_k = \begin{bmatrix}
				h_k \\
				h_ku_k
			\end{bmatrix} \\
			\textbf{F}_k = \begin{bmatrix} \label{eq: 2-layer SWE 2}
				h_ku_k \\
				\frac{(h_ku_k)^2}{h_k} + \frac{1}{2}gh_k^2 + gh_{k_{(+)}}h_k \\
			\end{bmatrix}
			= \begin{bmatrix}
				q_k \\
				\sigma_k
			\end{bmatrix} \\
			\textbf{S}_{k,0} = \begin{bmatrix}
				0 \\
				-R_{k+\frac{1}{2}}+R_{k-\frac{1}{2}} \\
			\end{bmatrix}
			= \begin{bmatrix}\label{eq: topographic source terms}
				0 \\
				g h_{k_{(+)}} \frac{\partial z_{k+1/2}}{\partial x} - g(h_{k_{(+)}} + h_k)\frac{\partial z_{k-1/2}}{\partial x}
			\end{bmatrix} \\
			\textbf{S}_{k,f} = \begin{bmatrix}
				0 \\
				\tau_{k+\frac{1}{2}} - \tau_{k-\frac{1}{2}}
			\end{bmatrix}
			= \begin{bmatrix}\label{eq: friction source terms}
				0 \\
				(1-\delta_{nk})\frac{2\nu(u_{k_{(+)}}-u_k)}{h_{k_{(+)}}+h_k} - \left((1-\delta_{1k}) \frac{2\nu(u_k-u_{k_{(-)}})}{h_k+h_{k_{(-)}}} - \delta_{1k} \frac{gn^2u_k|u_k|}{\sqrt[3]{H}}\right)
			\end{bmatrix}
		\end{gather}
		Where $k$ refers to the index of the layer under consideration, labelled in ascending order from layer $1$ at the bed, to layer $n$ at the free surface. $k+1/2$ and $k-1/2$ refer respectively to the upper and lower interface for layer $k$. The subscript $k_{(+)}$ refers to the properties of the flow above layer $k$ and the subscript $k_{(-)}$ refers to the properties of the flow below layer $k$, which are defined respectively as:
		\begin{gather}
			h_{k_{(+)}} = \sum_{k=k+1}^{n} h_k \quad \textrm { , } \quad h_{k_{(-)}} = \sum_{k=1}^{k-1} h_k \notag \\
			u_{k_{(+)}} = \frac{\sum_{k=k+1}^{n} h_ku_k}{h_{k_{(+)}}} \quad \textrm { , } \quad u_{k_{(-)}} = \frac{\sum_{k=1}^{k-1} h_ku_k}{h_{k_{(-)}}}
		\end{gather}
		$R_{k+1/2}$ and $R_{k-1/2}$ refer to the reaction forces exerted at the interfaces between the layers, with $R_{k+1/2}$ denoting the reaction force of layer $k$ onto the fluid above and $R_{k-1/2}$ denoting the reaction force exerted on layer $k$ by the fluid or bed beneath it. $\tau_{k+1/2}$ and $\tau_{k-1/2}$ represent the interlayer viscous friction effect induced at the upper and lower interfaces of layer $k$. The interlayer friction terms are derived for a multi-layer cell by applying a finite difference approximation, across the depth of the fluid layer $k$, to the viscous stress component of the incompressible Navier-Stokes system, as proposed by Audusse et al., \cite{RN377}:
		\begin{equation}\label{eq: interlayer friction term}
			\int_{z_{k-\frac{1}{2}}}^{z_{k+\frac{1}{2}}} \frac{\partial}{\partial z} \left( \nu \frac{\partial u}{\partial z}\right)dz 
			= \nu \frac{\partial u}{\partial z}\Big|_{z_{k+\frac{1}{2}}} -  \nu \frac{\partial u}{\partial z}\Big|_{z_{k-\frac{1}{2}}} 
			\approx \frac{2\nu(u_{k_{(+)}}-u_{k})}{h_{k_{(+)}}+h_k} - \frac{2\nu(u_{k}-u_{k_{(-)}})}{h_{k}+h_{k_{(-)}}} 
			= \tau_{k+\frac{1}{2}} - \tau_{k-\frac{1}{2}}
		\end{equation}   
		For the case where $k=1$, considering the layer which flows over the bed, $\tau_{k-1/2} = \tau_0$ which is instead derived from Manning's equation (\ref{eq: Manning's Friction}), where $H$ is the total depth of flow for the whole structure cell. The particular form of the viscous effect on the base of the fluid layer, $\tau_{k-1/2}$, is accounted for by Kronecker delta in (\ref{eq: friction source terms}), which is defined as:
		\begin{equation}\label{eq: delta}
			\delta_{\alpha k} = \begin{dcases}
				1 \textrm{ if } k = \alpha \\
				0 \textrm{ if } k \neq \alpha
			\end{dcases}
		\end{equation}
		The Kronecker delta also ensures that the $\tau_{k+1/2}$ term is zero at the free surface for layer $n$. The source terms for structure cells are also illustrated in Figure \ref{fig: Structure Cells Source Terms}. Effects relating to stresses as a result of volumetric deformation are not considered necessary to include due to their minor influence \cite{RN378}. For simplicity, wind friction effects on the free surface are also ignored however, wind friction effects can be easily added should the required wind data be available and the effects deemed necessary to include. \\
		\begin{figure}[hbt!]
			\centering
			\begin{tikzpicture}
				\draw[thick] (12.5,0) -- (12.5,0.01);
				\draw[thick] (3.5,0) -- (8.5,0); 
				\draw[dashed] (3,0) -- (3,5.25) node[above] {$II$};
				\draw[dashed] (7,0) -- (7,5.25) node[above] {$SI$};
				\draw[dashed] (11,0) -- (11,5.25) node[above] {$II$};
				
				\fill[fill=blue!35!white, draw=black] (3,0) rectangle (7,1.5); 
				\fill[fill=blue!30!white, draw=black] (3,1.5) rectangle (7,3.5); 
				\fill[fill=blue!25!white, draw=black] (3,3.5) rectangle (7,5); 
				\fill[fill=blue!35!white, draw=black] (7,-0.5) rectangle (11,1.5); 
				\fill[fill=blue!30!white, draw=black] (7,1.5) rectangle (11,3.5); 
				\fill[fill=blue!25!white, draw=black] (7,3.5) rectangle (11,4.75); 
				\fill[fill=black!25!white, draw=black] (6.8,1.5) rectangle (7.2,3.5); 
				
				\draw[thick,<->] (2.5,0) -- (2.5,1.5);
				\node[left] at (2.4, 0.75) {\large$h_{i,1}$};
				\draw[thick,<->] (2.5,1.5) -- (2.5,3.5);
				\node[left] at (2.4, 2.5) {\large$h_{i,2}$};
				\draw[thick,<->] (2.5,3.5) -- (2.5,5);
				\node[left] at (2.4, 4.25) {\large$h_{i,3}$};
				\draw[thick,<->] (11.5,-0.5) -- (11.5,1.5);
				\node[right] at (11.6, 0.5) {\large$h_{i+1,1}$};
				\draw[thick,<->] (11.5,1.5) -- (11.5,3.5);
				\node[right] at (11.6, 2.5) {\large$h_{i+1,2}$};
				\draw[thick,<->] (11.5,3.5) -- (11.5,5);
				\node[right] at (11.6, 4.25) {\large$h_{i+1,3}$};
				
				\draw[thick] (8.5,1.6) -- (9.5,1.6);
				\draw[thick] (9.5,1.6) -- (9.25,1.7);
				\node[right] at (9.5,1.695) {\large$\tau_{i+1,\frac{3}{2}}$};
				\draw[thick, ->] (7.75,1.5) -- (7.75,2.25) node[above] {$R_{i+1,\frac{3}{2}}$};
				\draw[thick] (8.5,1.4) -- (9.5,1.4);
				\draw[thick] (8.5,1.4) -- (8.75,1.3);
				\draw[thick] (8.5,3.6) -- (9.5,3.6);
				\draw[thick] (9.5,3.6) -- (9.25,3.7);
				\node[right] at (9.5,3.75) {\large$\tau_{i+1,\frac{5}{2}}$};
				\draw[thick, ->] (7.55,3.5) -- (7.55,4) node[above] {$R_{i+1,\frac{5}{2}}$};
				\draw[thick] (8.5,3.4) -- (9.5,3.4);
				\draw[thick] (8.5,3.4) -- (8.75,3.3);
				\draw[thick] (8.5,-0.4) -- (9.5,-0.4);
				\draw[thick] (9.5,-0.4) -- (9.25,-0.3);
				\node[right] at (9.5,-0.35) {\large$\tau_{i+1,b}$};
				\draw[thick, ->] (7.55,-0.5) -- (7.55,0.5) node[above] {$R_{i+1,b}$};
				\draw[thick] (4.5,1.6) -- (5.5,1.6);
				\draw[thick] (5.5,1.6) -- (5.25,1.7);
				\node[right] at (5.5,1.695) {\large$\tau_{i,\frac{3}{2}}$};
				\draw[thick, ->] (3.55,1.5) -- (3.55,2.25) node[above] {$R_{i,\frac{3}{2}}$};
				\draw[thick] (4.5,1.4) -- (5.5,1.4);
				\draw[thick] (4.5,1.4) -- (4.75,1.3);
				\draw[thick] (4.5,3.6) -- (5.5,3.6);
				\draw[thick] (5.5,3.6) -- (5.25,3.7);
				\node[right] at (5.5,3.75) {\large$\tau_{i,\frac{5}{2}}$};
				\draw[thick, ->] (3.55,3.5) -- (3.55,4) node[above] {$R_{i,\frac{5}{2}}$};
				\draw[thick] (4.5,3.4) -- (5.5,3.4);
				\draw[thick] (4.5,3.4) -- (4.75,3.3);
				\draw[thick] (4.5,0.1) -- (5.5,0.1);
				\draw[thick] (5.5,0.1) -- (5.25,0.2);
				\node[right] at (5.5,0.15) {\large$\tau_{i,b}$};
				\draw[thick, ->] (3.55,0) -- (3.55,0.6) node[above] {$R_{i,b}$};
				
				\draw[thick, dashed] (8, -0.5) -- (12, -0.5) node[right] {\large$z_{i+1,b}$};
				\draw[thick, dashed] (2, 1.5) -- (12, 1.5) node[right] {\large$z_{i+1,\frac{3}{2}}$};
				\draw[thick, dashed] (2, 3.5) -- (12, 3.5) node[right] {\large$z_{i+1,\frac{5}{2}}$};
				\draw[thick, dashed] (7, 0) -- (2, 0) node[left] {\large$z_{i,b}$};
				\node[left] at (2,1.5) {\large$z_{i,\frac{3}{2}}$};
				\node[left] at (2,3.5) {\large$z_{i,\frac{5}{2}}$};
				
				\node[below] at (5,-0.25) {\LARGE $\textbf{U}_{i}$};
				\node[below] at (9,-0.75) {\LARGE $\textbf{U}_{i+1}$};
				
				\node[] at (5,0.75) {\large $\textbf{U}_{i,1}$};
				\node[] at (9,0.5) {\large $\textbf{U}_{i+1,1}$};
				\node[] at (5,2.5) {\large $\textbf{U}_{i,2}$};
				\node[] at (9,2.5) {\large $\textbf{U}_{i+1,2}$};
				\node[] at (5,4.25) {\large $\textbf{U}_{i,3}$};
				\node[] at (9,4.125) {\large $\textbf{U}_{i+1,3}$};
				
			\end{tikzpicture}
			\caption{Annotation of the source terms for example structure cells and their component sub-cells on uneven bed topography. $R$ represents a reaction force induced as a result of the uneven bed topography, $\tau$ represents a friction force acting at a layer interface, $z$ denotes the elevation above the bed and $h$ denotes the water depth in the sub-cell. $\textbf{U}_i$ is the vector of conserved variables for the $i$th whole cell, which is equal to the sum of the conserved variables for the component sub cells $\textbf{U}_{i,k}$.}
			\label{fig: Structure Cells Source Terms}
		\end{figure}
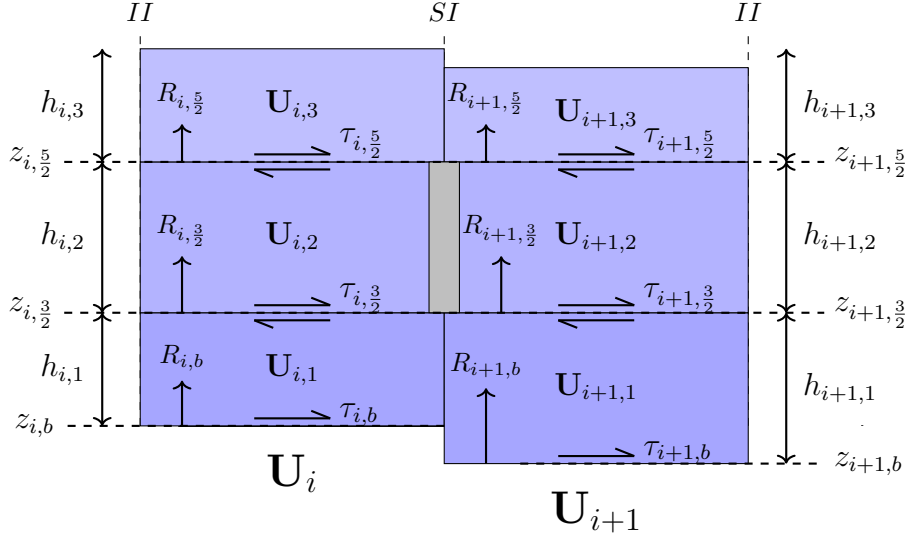 \\
		The domain is divided into cells $(\textbf{V}_i)_{i \in \mathbb{Z}}$ and the discretised first order finite volume scheme is given by:
		\begin{equation}\label{eq: Conservative Update}
			\textbf{U}^{n+1}_i = \textbf{U}^{n}_i - \frac{\Delta t}{\Delta x}\left[\textbf{F}_{i+\frac{1}{2}} -  \textbf{F}_{i-\frac{1}{2}}\right] + \Delta t \textbf{S}\left(\textbf{U}_i^n \right)
		\end{equation}
		Where the subscript $i$ represents the $i$th cell, the superscript $n$ represents the $n$th time level and $\Delta x$ and $\Delta t$ represent the cell size and time step respectively. $\textbf{F}_{i-1/2}$ and $\textbf{F}_{i+1/2}$ represent the numerical fluxes at the $i\pm1/2$ interfaces respectively. For the structure cells, it is the constituent sub-cells which are updated using the following modification of (\ref{eq: Conservative Update}):
		\begin{equation}\label{eq: Sub-Cell Conservative Update}
			\textbf{U}^{n+1}_{i,k} = \textbf{U}^{n}_{i,k} - \frac{\Delta t}{\Delta x}\left[\textbf{F}_{i+\frac{1}{2},k} -  \textbf{F}_{i-\frac{1}{2},k}\right] + \Delta t \textbf{S}\left(\textbf{U}_{i,k}^n \right)
		\end{equation}
		Where $\textbf{U}^{n}_{i,k}$ represents the conserved variables for the $k$th sub-cell in the $i$th structure cell at time level $n$. $\textbf{F}_{i-1/2, k}$ and $\textbf{F}_{i+1/2, k}$ represent the numerical fluxes at the $k$th layer of the $i\pm1/2$ interfaces respectively. Although a 1D scheme is implemented in this case, implementation as a 2D scheme requires no fundamental changes to the method. 
		
		\subsection{Numerical Flux Computation}
		The process for resolving fluxes is dependent on the type of interface (NI, II or SI). For structure and intermediate interfaces Harten-Lax-van Leer (HLL) approximate Riemann solvers \cite{RN269} are used to resolve the intercell numerical fluxes. For normal interfaces, other suitable approximate Riemann solvers may be used, however, HLL approximate Riemann solvers are recommended for consistency. 
		
		\subsubsection{Normal Interfaces}
		\begin{figure}[hbt!]
			\centering
			\begin{tikzpicture}
				\node[below] at (2, -0.5) {(\textbf{a})};
				\draw[thick] (-0.5,0) -- (4.5,0);
				\fill[fill=blue!25!white, draw=black] (0,0) rectangle (2,2); 
				\fill[fill=blue!25!white, draw=black] (2,0) rectangle (4,1.75); 
				\draw[thick, dashed] (0,0) -- (0,2.25);
				\draw[thick, dashed] (2,0) -- (2,2.25) node[above] {$NI$};
				\draw[thick, dashed] (4,0) -- (4,2.25);
				\draw[->, thick] (1.6,0.85) -- (2.4,0.85); 
				\draw[->, thick] (2.4,0.65) -- (1.6,0.65); 
				\node[above] at (2, 0.9) {$\textbf{F}_{i-\frac{5}{2}}$};
				\fill[fill=black, draw=black] (1,0) circle (0.1cm);
				\node[below] at (1,-0.1) {$x_{i-3}$};
				\fill[fill=black, draw=black] (3,0) circle (0.1cm);
				\node[below] at (3,-0.1) {$x_{i-2}$};
				
				\draw[thick] (5.5,0) -- (8.5, 0);
				\node [] (A) at ( 7, 0) {};
				\node [] (B) at (5.75, 2) {};
				\node [] (C) at (8.25, 2) {};
				\draw (A) -- (B) -- (C) -- (A);
				\begin{scope} 
					\fill [blue!10!white] (A.center) -- (B.center) -- (C.center) -- cycle;
				\end{scope}
				\draw[thick,->] (5.5,0) -- (8.5,0) node[right] {x}; 
				\node[below] at (7, -0.5) {(\textbf{b})};
				\draw[thick,->] (7,0) -- (7,2.25) node[above] {t};
				\node[below] at (7,0) {0};
				\draw[thick] (7,0) -- (5.65,2);
				\draw[thick] (7,0) -- (5.75,2) node[left] {$S^-$};
				\draw[thick] (7,0) -- (8.25,2) node[right] {$S^+$};
				\draw[thick] (7,0) -- (8.35,2);
				\node[above] at (5.5,1) {$h_{i-3}$};
				\node[below] at (5.5,1) {$u_{i-3}$};
				\node[above] at (8.5,1) {$h_{i-2}$};
				\node[below] at (8.5,1) {$u_{i-2}$};
				\node[above] at (6.65,1.15) {$h_*$};
				\node[above] at (7.35,1.15) {$u_*$};
			\end{tikzpicture}
			\caption{(a) Example normal interface with adjacent normal cells and (b) the general structure of the general solution of the Riemann problem for a normal interface. $S^-$ is the left wave speed and $S^+$ is the right wave speed, as defined in Algorithm \ref{Alg: Wavespeeds}. $h_*$ and $u_*$ denote the conserved variables in the star region. $\textbf{F}_{i-\frac{5}{2}}$ denotes the numerical flux at the interface.}
			\label{fig: Normal Interface}
		\end{figure}
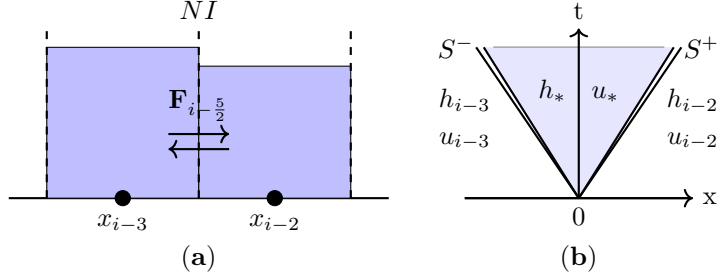
		A robust algorithm presented by Glenis et al. \cite{RN379} is used to calculate wave speeds for the Riemann problem, which is outlined in Algorithm \ref{Alg: Wavespeeds}. Following calculation of the wavespeeds, a standard HLL approximate Riemann solver (\ref{eq: HLL Solver}) is used to determine numerical fluxes across the normal interface.
		\begin{equation}\label{eq: HLL Solver} \textbf{F}_{i+\frac{1}{2}} = \begin{cases}
				\textbf{F}_{i} \textrm{ if } S^- > 0 \\[6pt]
				\textbf{F}^{hll} = \frac{S^+\textbf{F}_{i}-S^-\textbf{F}_{i+1}+S^+S^-(\textbf{U}_{i+1}-\textbf{U}_i)}{S^+-S^-} \textrm{ if } S^- \leq 0 \leq S^+ \\[6pt]
				\textbf{F}_{i+1} \textrm{ if } S^+ < 0
		\end{cases}  \end{equation}
		As discussed prior, other suitable approximate Riemann solvers may also be used however, use of a HLL solver is recommended for consistency.
		\begin{algorithm}
			\caption{Calculation of wavespeeds \cite{RN379}. An initial approximation ($h_0$) of the depth in the star region ($h_*$) using a two-rarefaction approximate state Riemann solver is used to determine whether a two-rarefaction or two-shock approximation is optimal. For a multi-layer system, the wave celerity is defined as $c_{i,k} = \sqrt{g(h_{i,k}+h_{i,k_{(+)}})}$, where $c_{i,k}$ is the celerity for cell $i$ layer $k$, $h_{i,k}$ is the thickness of cell $i$, layer $k$ and $h_{i,k_{(+)}}$ is the depth of water in cell $i$ above layer $k$.}\label{Alg: Wavespeeds}
			$g \leftarrow 9.81$ms$^{-2}$ \\
			\If{$h_i \wedge h_{i+1} > 0$}{\Comment{Initial two-rarefaction approximation}
				\begin{gather*}
					c_i \leftarrow \sqrt{gh_i} \quad \textrm{ , } \quad c_{i+1} \leftarrow \sqrt{gh_{i+1}} \\
					h_0 \leftarrow \frac{1}{g}\left(\frac{1}{2}(c_i+c_{i+1}) + \frac{1}{4}(u_i-u_{i+1}) \right)^2
				\end{gather*}
				\uIf{$h_0 \leq \textrm{min}(h_i, h_{i+1})$}{\Comment{Use two-rarefaction approximate state Riemann solver}
				$$h_* \leftarrow h_0 $$
				}
			\uElseIf{$h_0 > \textrm{min}(h_i, h_{i+1})$}{\Comment{Use two-shock approximate state Riemann solver}
				\begin{gather*}
					p_i \leftarrow \sqrt{\frac{g(h_0+h_i)}{2h_0h_i}} \quad \textrm{ , } \quad p_{i+1} \leftarrow \sqrt{\frac{g(h_0+h_{i+1})}{2h_0h_{i+1}}} \\
					h_* \leftarrow \frac{p_ih_i + p_{i+1}h_{i+1} + u_i - u_{i+1}}{p_i + p_{i+1}} 
				\end{gather*}
				}
			}
			\begin{gather*}
				\alpha_i \leftarrow \begin{dcases}
					\frac{\sqrt{0.5(h_*+h_i)h_*}}{h_i} \textrm{ if } h_* > h_i \\
					\quad \qquad \ \ 1 \qquad \quad \ \ \textrm{ if } h_* \leq h_i
				\end{dcases} \quad \textrm{ , } \quad
				\alpha_{i+1} \leftarrow \begin{dcases}
					\frac{\sqrt{0.5(h_*+h_{i+1})h_*}}{h_{i+1}} \textrm{ if } h_* > h_{i+1} \\
					\quad \qquad \ \ 1 \qquad \quad \ \ \textrm{ if } h_* \leq h_{i+1}
				\end{dcases} \\
				S^- \leftarrow u_i - \alpha_ic_i \quad \textrm{ , } \quad S^+ \leftarrow u_{i+1} + \alpha_{i+1}c_{i+1}
			\end{gather*}
			\ElseIf{$h_i = 0 \wedge h_{i+1} > 0$}{\Comment{Left dry bed}
			\begin{gather*}
				S^- \leftarrow u_{i+1} - 2c_{i+1} \quad \textrm{ , } \quad S^+ \leftarrow u_{i+1} + c_{i+1}
			\end{gather*}
			\uElseIf{$h_{i+1} = 0 \wedge h_i > 0$}{\Comment{Right dry bed}
			\begin{gather*}
				S^- \leftarrow u_i - c_i \quad \textrm{ , } \quad S^+ \leftarrow u_i + 2c_{i}
			\end{gather*}
			}
			}
		\end{algorithm}
		
		\subsubsection{Structure Interfaces}
		At a structure interface the layers of flow and can be divided into \textit{open} and \textit{closed} as shown in Figure \ref{fig: Open and Closed Layers}.
		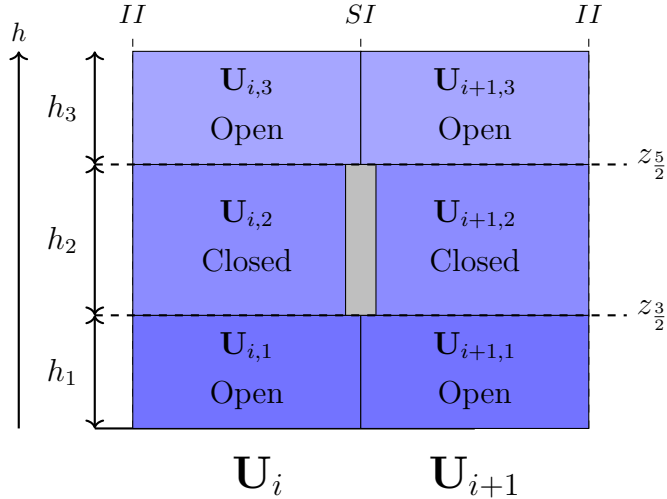
\begin{figure}[hbt!]
			\centering
			\begin{tikzpicture}
				\draw[thick, ->] (2.5,0) -- (2.5,5) node[above] {$h$};
				\draw[thick] (11.5,0) -- (11.5,0.01);
				\draw[thick] (3.5,0) -- (8.5,0); 
				\draw[dashed] (4,0) -- (4,5.25) node[above] {$II$};
				\draw[dashed] (7,0) -- (7,5.25) node[above] {$SI$};
				\draw[dashed] (10,0) -- (10,5.25) node[above] {$II$};
				
				\fill[fill=blue!55!white, draw=black] (4,0) rectangle (7,1.5); 
				\fill[fill=blue!45!white, draw=black] (4,1.5) rectangle (7,3.5); 
				\fill[fill=blue!35!white, draw=black] (4,3.5) rectangle (7,5); 
				\fill[fill=blue!55!white, draw=black] (7,0) rectangle (10,1.5); 
				\fill[fill=blue!45!white, draw=black] (7,1.5) rectangle (10,3.5); 
				\fill[fill=blue!35!white, draw=black] (7,3.5) rectangle (10,5); 
				\fill[fill=black!25!white, draw=black] (6.8,1.5) rectangle (7.2,3.5); 
				
				\draw[thick,<->] (3.5,0) -- (3.5,1.5);
				\node[left] at (3.4, 0.75) {\large$h_1$};
				\draw[thick,<->] (3.5,1.5) -- (3.5,3.5);
				\node[left] at (3.4, 2.5) {\large$h_2$};
				\draw[thick,<->] (3.5,3.5) -- (3.5,5);
				\node[left] at (3.4, 4.25) {\large$h_3$};
				
				\draw[thick, dashed] (3.5, 1.5) -- (10.5, 1.5) node[right] {\large$z_{\frac{3}{2}}$};
				\draw[thick, dashed] (3.5, 3.5) -- (10.5, 3.5) node[right] {\large$z_{\frac{5}{2}}$};
				
				\node[below] at (5.65,-0.25) {\LARGE $\textbf{U}_{i}$};
				\node[below] at (8.5,-0.25) {\LARGE $\textbf{U}_{i+1}$};
				
				\node[above] at (5.5,0.75) {\large $\textbf{U}_{i,1}$};
				\node[below] at (5.5,0.75) {\large Open};
				\node[above] at (8.5,0.75) {\large $\textbf{U}_{i+1,1}$};
				\node[below] at (8.5,0.75) {\large Open};
				\node[above] at (5.5,2.5) {\large $\textbf{U}_{i,2}$};
				\node[below] at (5.5,2.5) {\large Closed};
				\node[above] at (8.5,2.5) {\large $\textbf{U}_{i+1,2}$};
				\node[below] at (8.5,2.5) {\large Closed};
				\node[above] at (5.5,4.25) {\large $\textbf{U}_{i,3}$};
				\node[below] at (5.5,4.25) {\large Open};
				\node[above] at (8.5,4.25) {\large $\textbf{U}_{i+1,3}$};
				\node[below] at (8.5,4.25) {\large Open};
				
			\end{tikzpicture}
			\caption{Designation of open and closed layers at a structure interface.}
			\label{fig: Open and Closed Layers}
		\end{figure}
		\textit{Open} layers are considered as having a transmissive boundary at the structure interface, with the portion of the structure interface shared by the adjacent sub-cells having no influence on the exchange of conserved variables. \textit{Closed} layers are considered as having a reflective boundary at the structure interface due to the presence of the structure. For each open layer, a single Riemann problem must be constructed and solved whereas, at each closed layer two Riemann problems must be constructed and solved, as shown in Figure \ref{fig: SI Fluxes}. Solution of two Riemann problems for a closed layer is necessary to implement the reflective boundary condition at the structure interface, which reflects the flow in both the left and right sub-cells. This process is based on the assumption that the vertical velocity of the flow is negligible, which is a fundamental assumption for the derivation of the shallow water equations, and therefore the direction of the flow can be considered to be primarily parallel to the bed. 
		
		The numerical flux for each layer is determined by applying (\ref{eq: 2-layer SWE 1}) to each layer, where the numerical flux for a layer is given as:
		\begin{gather}\label{eq: Submerged Flux}
			\textbf{F}_k = \begin{bmatrix}
				h_ku_k \\
				\frac{(h_ku_k)^2}{h_k} + \frac{1}{2}gh_k^2 + gh_{k_{(+)}}h_k \\
			\end{bmatrix}
			= \begin{bmatrix}
				q_k \\
				\sigma_k 
			\end{bmatrix}
		\end{gather}
		Which can then be used to determine the flux at the interface using a standard HLL approximate Riemann solver (\ref{eq: HLL Solver}).
		\begin{figure*}[t!]
			\centering
			\begin{subfigure}[t]{0.5\textwidth}
				\centering
				\begin{tikzpicture}
					\draw[dashed] (4,0) -- (4,7);
					\draw[dashed] (7,0) -- (7,7) ;
					\draw[dashed] (10,0) -- (10,7);
					
					\fill[fill=blue!55!white, draw=black] (4,0) rectangle (7,1.5); 
					\fill[fill=blue!55!white, draw=black] (7,0) rectangle (10,1.5); 
					\node[below] at (5.5,0) {\large $\textbf{U}_{i,1}$};
					\node[below] at (8.5,0) {\large $\textbf{U}_{i+1,1}$};
					\node[above] at (5.5, 0.75) {\large$h_{i,1}$};
					\node[below] at (5.5, 0.75) {\large$u_{i,1}$};
					\node[above] at (8.5, 0.75) {\large$h_{i+1,2}$};
					\node[below] at (8.5, 0.75) {\large$u_{i+1,1}$};
					\draw[->, thick] (6.65,0.7) -- (7.35,0.7); 
					\draw[->, thick] (7.35,0.5) -- (6.65,0.5); 
					\node[above] at (7, 0.725) {\large$\textbf{F}_{i+\frac{1}{2},1}$};
					
					\fill[fill=blue!45!white, draw=black] (4,2.75) rectangle (7,4.75); 
					\fill[fill=blue!45!white, draw=black] (7,2.75) rectangle (10,4.75); 
					\fill[fill=black!25!white, draw=black] (6.8,2.75) rectangle (7.2,4.75); 
					\node[below] at (5.5,2.75) {\large $\textbf{U}_{i,2}$};
					\node[below] at (8.5,2.75) {\large $\textbf{U}_{i+1,2}$};
					\node[above] at (4.75, 3.75) {\large$h_{i,2}$};
					\node[below] at (4.75, 3.75) {\large$u_{i,2}$};
					\draw[->, thick] (6,3.85) -- (6.7,3.85); 
					\draw[->, thick] (6.7,3.65) -- (6,3.65); 
					\node[above] at (9.25, 3.75) {\large$h_{i+1,2}$};
					\node[below] at (9.25, 3.75) {\large$u_{i+1,2}$};
					\draw[->, thick] (7.3,3.85) -- (8,3.85); 
					\draw[->, thick] (8,3.65) -- (7.3,3.65); 
					\node[above] at (6.1, 3.95) {\large$\textbf{F}_{i+\frac{1}{2},2_L}$};
					\node[above] at (7.9, 3.95) {\large$\textbf{F}_{i+\frac{1}{2},2_R}$};
					
					\fill[fill=blue!35!white, draw=black] (4,6) rectangle (7,7.5); 
					\fill[fill=blue!35!white, draw=black] (7,6) rectangle (10,7.5); 
					\node[below] at (5.5,6) {\large $\textbf{U}_{i,3}$};
					\node[below] at (8.5,6) {\large $\textbf{U}_{i+1,3}$};
					\node[above] at (5.5, 6.75) {\large$h_{i,3}$};
					\node[below] at (5.5, 6.75) {\large$u_{i,3}$};
					\node[above] at (8.5, 6.75) {\large$h_{i+1,3}$};
					\node[below] at (8.5, 6.75) {\large$u_{i+1,3}$};
					\draw[->, thick] (6.65,6.65) -- (7.35,6.65); 
					\draw[->, thick] (7.35,6.45) -- (6.65,6.45); 
					\node[above] at (7, 6.725) {\large$\textbf{F}_{i+\frac{1}{2},3}$};
				\end{tikzpicture}
				\caption{Division of the structure cells into sub-cells and their respective properties. The subscripts $L$ and $R$ are used to differentiate between the left and right face of the structure interface.}
				\label{fig: SI Split}
			\end{subfigure}%
			~
			\begin{subfigure}[t]{0.5\textwidth}
				\centering
				\begin{tikzpicture}
				\draw[thick] (0,0) -- (3, 0);
				\node [] (A) at (1.5, 0) {};
				\node [] (B) at (0.25, 2) {};
				\node [] (C) at (2.75, 2) {};
				\draw (A) -- (B) -- (C) -- (A);
				\begin{scope} 
					\fill [blue!10!white] (A.center) -- (B.center) -- (C.center) -- cycle;
				\end{scope}
				\draw[thick,->] (0,0) -- (3,0) node[right] {x}; 
				\draw[thick,->] (1.5,0) -- (1.5,2.25) node[above] {t};
				\node[below] at (1.5,0) {0};
				\draw[thick] (1.5,0) -- (0.15,2);
				\draw[thick] (1.5,0) -- (0.25,2) node[left] {$S_{1}^-$};
				\draw[thick] (1.5,0) -- (2.75,2) node[right] {$S_{1}^+$};
				\draw[thick] (1.5,0) -- (2.85,2);
				\node[above] at (0.25,0.85) {$h_{i,1}$};
				\node[below] at (0.25,0.85) {$u_{i,1}$};
				\node[above] at (2.85,0.85) {$h_{i+1,1}$};
				\node[below] at (2.85,0.85) {$u_{i+1,1}$};
				\node[above] at (1.15,1.15) {$h_{*,1}$};
				\node[above] at (1.85,1.15) {$u_{*,1}$};
				
				\fill[fill=black!15!white, draw=none] (-0.5,3) rectangle (1.25,5);
				\draw[thick] (-2,3) -- (1, 3);
				\node [] (A) at (-0.5, 3) {};
				\node [] (B) at (-1.75, 5) {};
				\node [] (C) at (0.75, 5) {};
				\draw (A) -- (B) -- (C) -- (A);
				\begin{scope} 
					\fill [blue!10!white] (A.center) -- (B.center) -- (C.center) -- cycle;
				\end{scope}
				\draw[thick,->] (-2,3) -- (1,3) node[right] {x}; 
				\draw[thick,->] (-0.5,3) -- (-0.5,5.25) node[above] {t};
				\node[below] at (-0.5,3) {0};
				\draw[thick] (-0.5,3) -- (-1.85,5);
				\draw[thick] (-0.5,3) -- (-1.75,5) node[above] {$S_{2_L}^-$};
				\draw[thick] (-0.5,3) -- (0.75,5) node[above] {$S_{2_L}^+$};
				\draw[thick] (-0.5,3) -- (0.85,5);
				\node[above] at (-1.75,3.85) {$h_{i,2}$};
				\node[below] at (-1.75,3.85) {$u_{i,2}$};
				\node[above] at (0.75,3.85) {$h_{i,2}$};
				\node[below] at (0.65,3.85) {$-u_{i,2}$};
				\node[above] at (-0.15,4.15) {$h_{*,2}$};
				\node[above] at (-0.85,4.15) {$u_{*,2}$};

				\fill[fill=black!15!white, draw=none] (1.75,3) rectangle (3.5,5);
				\draw[thick] (2,3) -- (5,3);
				\node [] (A) at (3.5, 3) {};
				\node [] (B) at (2.25, 5) {};
				\node [] (C) at (4.75, 5) {};
				\draw (A) -- (B) -- (C) -- (A);
				\begin{scope} 
					\fill [blue!10!white] (A.center) -- (B.center) -- (C.center) -- cycle;
				\end{scope}
				\draw[thick,->] (2,3) -- (5,3) node[right] {x}; 
				\draw[thick,->] (3.5,3) -- (3.5,5.25) node[above] {t};
				\node[below] at (3.5,3) {0};
				\draw[thick] (3.5,3) -- (2.15,5);
				\draw[thick] (3.5,3) -- (2.25,5) node[above] {$S_{2_R}^-$};
				\draw[thick] (3.5,3) -- (4.75,5) node[above] {$S_{2_R}^+$};
				\draw[thick] (3.5,3) -- (4.85,5);
				\node[above] at (2.25,3.85) {$h_{i+1,2}$};
				\node[below] at (2.1,3.85) {$-u_{i+1,2}$};
				\node[above] at (4.85,3.85) {$h_{i+1,2}$};
				\node[below] at (4.85,3.85) {$u_{i+1,2}$};
				\node[above] at (3.15,4.15) {$h_{*,2}$};
				\node[above] at (3.85,4.15) {$u_{*,2}$};
				
				\draw[thick] (0,6) -- (3, 6);
				\node [] (A) at (1.5, 6) {};
				\node [] (B) at (0.25, 8) {};
				\node [] (C) at (2.75, 8) {};
				\draw (A) -- (B) -- (C) -- (A);
				\begin{scope} 
					\fill [blue!10!white] (A.center) -- (B.center) -- (C.center) -- cycle;
				\end{scope}
				\draw[thick,->] (0,6) -- (3,6) node[right] {x}; 
				\draw[thick,->] (1.5,6) -- (1.5,8.25) node[above] {t};
				\node[below] at (1.5,6) {0};
				\draw[thick] (1.5,6) -- (0.15,8);
				\draw[thick] (1.5,6) -- (0.25,8) node[left] {$S_{3}^-$};
				\draw[thick] (1.5,6) -- (2.75,8) node[right] {$S_{3}^+$};
				\draw[thick] (1.5,6) -- (2.85,8);
				\node[above] at (0.25,6.85) {$h_{i,3}$};
				\node[below] at (0.25,6.85) {$u_{i,3}$};
				\node[above] at (2.85,6.85) {$h_{i+1,3}$};
				\node[below] at (2.85,6.85) {$u_{i+1,3}$};
				\node[above] at (1.15,7.15) {$h_{*,3}$};
				\node[above] at (1.85,7.15) {$u_{*,3}$};
			\end{tikzpicture}
			\caption{The general structure of the general solution of the Riemann problems for an example structure interface shown in (a). The introduction of fictitious ghost cells for the purpose of implementing reflective boundary conditions are denoted by grey shading.}
			\label{fig: SI Riemann Problems}
			\end{subfigure}
			\caption{Method for resolving fluxes for the sub-cells adjacent to a structure interface.}
			\label{fig: SI Fluxes}
		\end{figure*}
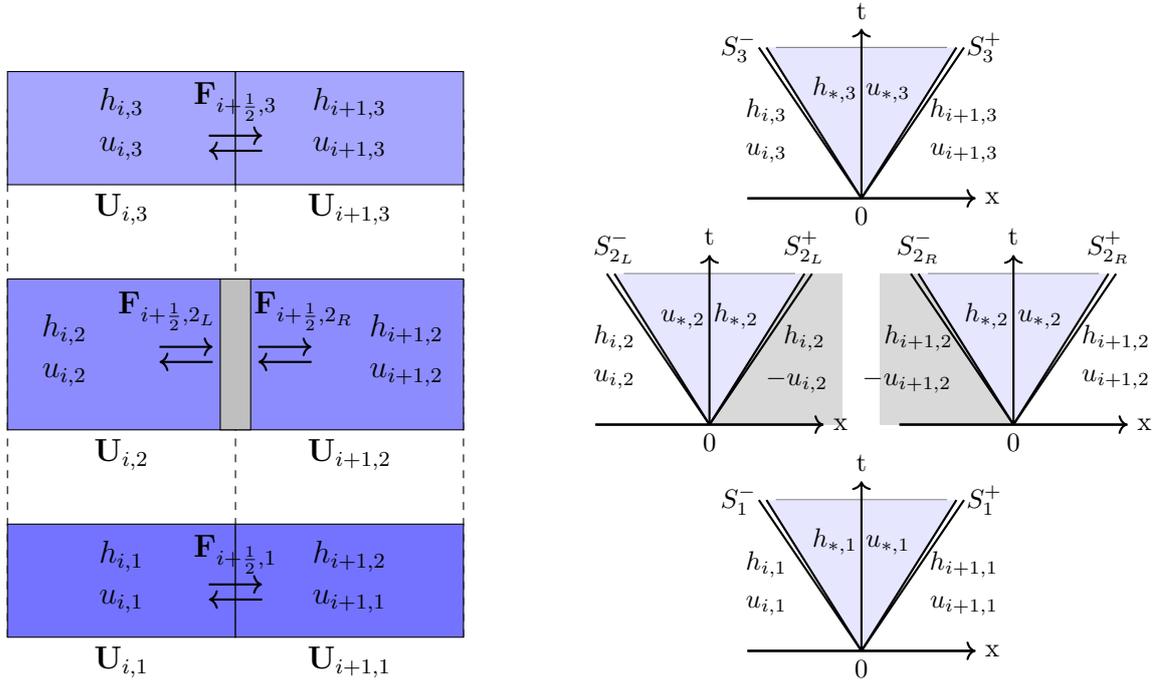
		The method for determining the fluxes at a structure interface is summarised in Algorithm \ref{Alg: SI Fluxes}.
		\begin{algorithm}
			\caption{Calculation of fluxes for an example structure interface as shown in Figure \ref{fig: SI Fluxes}. $k$ is the index of the layer under consideration, $n$ is the total number of layers at the structure interface.}\label{Alg: SI Fluxes}
			$g \leftarrow 9.81$ms$^{-2}$ \\
			$k \leftarrow 1$ \Comment{For the open layers} \\
			\While{$k \leq n$}{
				calculate $S^-_k$ , $S^+_k$ using Algorithm $(1)$ \Comment{Calculate wavespeeds} \\
				\Comment{Calculate layer flux}
				\begin{equation*} \textbf{F}_{i,k} \leftarrow \begin{bmatrix}
						h_{i,k}u_{i,k} \\
						\frac{q_{i,k}^2}{h_{i,k}} + \frac{1}{2}gh_{i,k}^2 + gh_{i,k_{(+)}}h_{i,k} \\ 
				\end{bmatrix} \textrm{ , } \textbf{F}_{i+1,k} \leftarrow \begin{bmatrix}
						h_{i+1,k}u_{i+1,k} \\
						\frac{q_{i+1,k}^2}{h_{i+1,k}} + \frac{1}{2}gh_{i+1,k}^2 + gh_{i+1,k_{(+)}}h_{i+1,k} \\ 
					\end{bmatrix}
				\end{equation*}
				\begin{equation*} \textbf{F}_{i+\frac{1}{2},k} \leftarrow \begin{cases}
						\textbf{F}_{i,k} \textrm{ if } S^-_k > 0 \\[6pt]
						\textbf{F}^{hll} = \frac{S^+\textbf{F}_{i,k}-S^-_k\textbf{F}_{i+1,k}+S^+_kS^-_k(\textbf{U}_{i+1,k}-\textbf{U}_{i,k})}{S^+_k-S^-_k} \textrm{ if } S^-_k \leq 0 \leq S^+_k \\[6pt] \tag{6}
						\textbf{F}_{i+1,k} \textrm{ if } S^+_k < 0
				\end{cases}  \end{equation*}
				$k \leftarrow k+2$ \Comment{Advance to next open layer} \\
			} 
			$k \leftarrow 2$ \Comment{For the closed layer} \\
			$h_{i+1,ghost} \leftarrow h_{i,k}$ \Comment{Right ghost cell water depth} \\
			$u_{i+1,ghost} \leftarrow -u_{i,k}$ \Comment{Right ghost cell water velocity} \\
			calculate $S^-_{k_L}$ , $S^+_{k_L}$ using Algorithm $(1)$ \Comment{Calculate wavespeeds} \\
			calculate $\textbf{F}_{i+\frac{1}{2},k_L}$ using $(6)$ \Comment{Flux for the left side of the structure} \newline \\ 
			$h_{i,ghost} \leftarrow h_{i,k}$ \Comment{Left ghost cell water depth} \\
			$u_{i,ghost} \leftarrow -u_{i,k}$ \Comment{Left ghost cell water velocity} \\
			calculate $S^-_{k_R}$ , $S^+_{k_R}$ using Algorithm $(1)$ \Comment{Calculate wavespeeds} \\
			calculate $\textbf{F}_{i+\frac{1}{2},k_R}$ using $(6)$ \Comment{Flux for the right side of the structure} \\
		\end{algorithm}
		\subsubsection{Intermediate Interfaces}
		 In order to resolve fluxes with the adjacent sub-cells it is necessary to temporarily define layer properties for the intermediate cell as shown in Figure \ref{fig: Intermediate Cells}.
		\begin{figure}[hbt!]
			\centering
			\begin{tikzpicture}
				\draw[thick, ->] (2.5,0) -- (2.5,5) node[above] {$h$};
				\draw[thick] (11.5,0) -- (11.5,0.01);
				\draw[thick] (3.5,0) -- (8.5,0); 
				\draw[dashed] (4,0) -- (4,5.25) node[above] {$NI$};
				\draw[dashed] (7,0) -- (7,5.25) node[above] {$II$};
				\draw[dashed] (10,0) -- (10,5.25) node[above] {$SI$};
				
				\fill[fill=blue!25!white, draw=black] (4,0) rectangle (7,5); 
				\fill[fill=blue!55!white, draw=black] (7,0) rectangle (10,1.5); 
				\fill[fill=blue!45!white, draw=black] (7,1.5) rectangle (10,3.5); 
				\fill[fill=blue!35!white, draw=black] (7,3.5) rectangle (10,5); 
				\fill[fill=black!25!white, draw=black] (9.8,1.5) rectangle (10.2,3.5); 
				
				\draw[thick,<->] (3.5,0) -- (3.5,1.5);
				\node[left] at (3.4, 0.75) {\large$h_1$};
				\draw[thick,<->] (3.5,1.5) -- (3.5,3.5);
				\node[left] at (3.4, 2.5) {\large$h_2$};
				\draw[thick,<->] (3.5,3.5) -- (3.5,5);
				\node[left] at (3.4, 4.25) {\large$h_3$};
				
				\draw[thick, dashed] (3.5, 1.5) -- (10.5, 1.5) node[right] {\large$z_{\frac{3}{2}}$};
				\draw[thick, dashed] (3.5, 3.5) -- (10.5, 3.5) node[right] {\large$z_{\frac{5}{2}}$};
				
				\node[below] at (5.65,-0.25) {\LARGE $\textbf{U}_{i-1}$};
				\node[below] at (8.5,-0.25) {\LARGE $\textbf{U}_{i}$};
				
				\node[] at (8.5,0.75) {\large $\textbf{U}_{i,1}$};
				\node[] at (8.5,2.5) {\large $\textbf{U}_{i,2}$};
				\node[] at (8.5,4.25) {\large $\textbf{U}_{i,3}$};
				\node[above] at (5.5,0.75) {$h_{i-1,1}$};
				\node[below] at (5.5,0.75) {$u_{i-1,1}$};
				\node[above] at (5.5,2.5) {$h_{i-1,2}$};
				\node[below] at (5.5,2.5) {$u_{i-1,2}$};
				\node[above] at (5.5,4.25) {$h_{i-1,3}$};
				\node[below] at (5.5,4.25) {$u_{i-1,3}$};
			\end{tikzpicture}
			\caption{Temporary division of an intermediate cell into layers in order to resolve fluxes at a intermediate interface. $u_{i-1,1} = u_{i-1,2} = u_{i-1,3} = u_{i-1}$ where $u_{i-1}$ represents the average velocity for the whole intermediate cell.}
			\label{fig: Intermediate Cells}
		\end{figure}
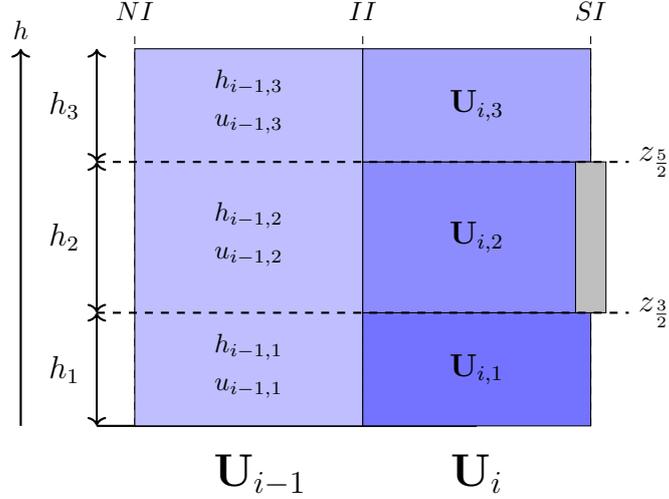
		The properties for the temporary layers in the intermediate interfaces are defined by assuming that the velocity in each layer is equal to the average velocity of the whole intermediate cell and that the depth in each layer is limited to the maximum depth capacity of the adjacent sub-cell. The fluxes for each layer can then be found using the process outlined for the open layers in Algorithm \ref{Alg: SI Fluxes}. \newpage
		\subsection{Conservative Updating of Conserved Variables}
		Once numerical fluxes have been resolved across all interfaces within the computational domain, the final procedure for each timestep is to update the conserved variables contained within each cell and sub-cell.
		\subsubsection{Normal Cells}
		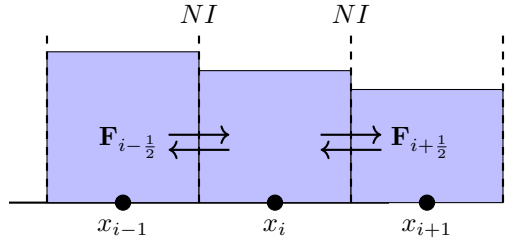
\begin{figure}[hbt!]
			\centering
			\begin{tikzpicture}
				\draw[thick] (-0.5,0) -- (4.5,0);
				\fill[fill=blue!25!white, draw=black] (0,0) rectangle (2,2); 
				\fill[fill=blue!25!white, draw=black] (2,0) rectangle (4,1.75); 
				\fill[fill=blue!25!white, draw=black] (4,0) rectangle (6,1.5); 
				\draw[thick, dashed] (0,0) -- (0,2.25);
				\draw[thick, dashed] (2,0) -- (2,2.25) node[above] {$NI$};
				\draw[thick, dashed] (4,0) -- (4,2.25) node[above] {$NI$};
				\draw[thick, dashed] (6,0) -- (6,2.25);
				\draw[->, thick] (1.6,0.9) -- (2.4,0.9); 
				\draw[->, thick] (2.4,0.7) -- (1.6,0.7); 
				\node[left] at (1.6, 0.8) {$\textbf{F}_{i-\frac{1}{2}}$};
				\fill[fill=black, draw=black] (1,0) circle (0.1cm);
				\node[below] at (1,-0.1) {$x_{i-1}$};
				\fill[fill=black, draw=black] (3,0) circle (0.1cm);
				\node[below] at (3,-0.1) {$x_{i}$};
				\draw[->, thick] (3.6,0.9) -- (4.4,0.9); 
				\draw[->, thick] (4.4,0.7) -- (3.6,0.7); 
				\node[right] at (4.4, 0.8) {$\textbf{F}_{i+\frac{1}{2}}$};
				\node[below] at (5,-0.1) {$x_{i+1}$};
				\fill[fill=black, draw=black] (5,0) circle (0.1cm);
			\end{tikzpicture}
			\caption{Illustration of the numerical fluxes at the normal interfaces bordering a normal cell.}
			\label{fig: Normal Cell Update}
		\end{figure}
		Normal cells are updated using equation (\ref{eq: Conservative Update}), which is standard for a one-dimensional Godunov type scheme. For cases involving variable bed topography, a well-balanced treatment of the topographic source terms can be achieved via the hydrostatic reconstruction method \cite{RN280} or via upwinding of the source terms \cite{RN281}. Suitable explicit or implicit treatment of the remaining source terms are both viable depending on the desired stability and admissible constraint of the stable timestep. For strong stability and the flexibility of advancing the solution at the timestep for the advection problem, the splitting method proposed by Liang and Marche \cite{RN27} is recommended:
		\begin{equation}\label{eq: Friction Source Term}
			q^{n+1}_i = q^{n}_i - \Delta t S^n_{i,c} = q^{n}_i - \Delta t \left(\frac{\tau_{i,f}}{1 + \Delta t\frac{\partial \tau_{i,f}}{\partial q_i}} \right)^n = q^{n}_i - \Delta t \left(\frac{C_iu_i|u_i|}{1 + \frac{2\Delta tC_{i,f}|q_i|}{h_i^2}} \right)^n 
		\end{equation}
		The following simple limiter is also recommended to ensure stability in regions where the water depth approaches zero:
		\begin{equation}
			S_{i,c}^n = \frac{q^n_i}{\Delta t} \textrm{ if } |\Delta t S_{i,c}^n| > |q_i^n|
		\end{equation}
		
		\subsubsection{Intermediate Cells}
		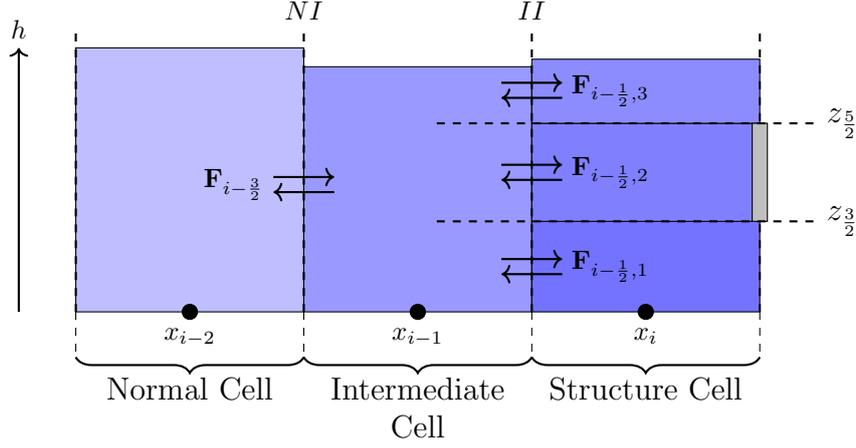
\begin{figure}[hbt!]
			\centering
			\begin{tikzpicture}
				\draw[thick, ->] (-0.75,0) -- (-0.75,3.5) node[above] {$h$};
				\fill[fill=blue!25!white, draw=black] (0,0) rectangle (3,3.5); 
				\fill[fill=blue!40!white, draw=black] (3,0) rectangle (6,3.25); 
				\fill[fill=blue!55!white, draw=black] (6,0) rectangle (9,1.2); 
				\fill[fill=blue!50!white, draw=black] (6,1.2) rectangle (9,2.5); 
				\fill[fill=blue!45!white, draw=black] (6,2.5) rectangle (9,3.35); 
				\draw[thick, dashed] (0,0) -- (0,3.75);
				\draw[thick, dashed] (3,0) -- (3,3.75) node[above] {$NI$};
				\draw[thick, dashed] (6,0) -- (6,3.75) node[above] {$II$};
				\draw[thick, dashed] (9,0) -- (9,3.75);
				\fill[fill=black!25!white, draw=black] (8.9,1.2) rectangle (9.1,2.5); 
				\draw[thick, dashed] (4.75, 1.2) -- (9.75, 1.2) node[right] {\large$z_{\frac{3}{2}}$};
				\draw[thick, dashed] (4.75, 2.5) -- (9.75, 2.5) node[right] {\large$z_{\frac{5}{2}}$};
				
				\draw[->, thick] (2.6,1.7875) -- (3.4,1.7875); 
				\draw[->, thick] (3.4,1.5875) -- (2.6,1.5875); 
				\node[left] at (2.6, 1.6875) {$\textbf{F}_{i-\frac{3}{2}}$};
				\draw[->, thick] (5.6,0.7) -- (6.4,0.7); 
				\draw[->, thick] (6.4,0.5) -- (5.6,0.5); 
				\node[right] at (6.4, 0.6) {$\textbf{F}_{i-\frac{1}{2},1}$};
				\draw[->, thick] (5.6,1.95) -- (6.4,1.95); 
				\draw[->, thick] (6.4,1.75) -- (5.6,1.75); 
				\node[right] at (6.4, 1.85) {$\textbf{F}_{i-\frac{1}{2},2}$};
				\draw[->, thick] (5.6,3.0375) -- (6.4,3.0375); 
				\draw[->, thick] (6.4,2.8375) -- (5.6,2.8375); 
				\node[right] at (6.4, 2.9375) {$\textbf{F}_{i-\frac{1}{2},3}$};
				\draw [decorate, decoration={brace,amplitude=6pt,raise=0pt}, thick] (9,-0.6) -- (6,-0.6); 
				\node[below] at (7.5,-0.75) {\large Structure Cell};
				\draw [decorate, decoration={brace,amplitude=6pt,raise=0pt}, thick] (6,-0.6) -- (3,-0.6); 
				\node[below] at (4.5,-0.75) {\large Intermediate};
				\node[below] at (4.5,-1.25) {\large Cell};
				\draw [decorate, decoration={brace,amplitude=6pt,raise=0pt}, thick] (3,-0.6) -- (0,-0.6); 
				\node[below] at (1.5,-0.75) {\large Normal Cell};
				\draw[dashed] (0,0) -- (0,-0.6);
				\fill[fill=black, draw=black] (1.5,0) circle (0.1cm);
				\node[below] at (1.5,-0.1) {$x_{i-2}$};
				\draw[dashed] (3,0) -- (3,-0.6);
				\fill[fill=black, draw=black] (4.5,0) circle (0.1cm);
				\node[below] at (4.5,-0.1) {$x_{i-1}$};
				\draw[dashed] (6,0) -- (6,-0.6);
				\fill[fill=black, draw=black] (7.5,0) circle (0.1cm);
				\node[below] at (7.5,-0.1) {$x_{i}$};
				\draw[dashed] (9,0) -- (9,-0.6);
			\end{tikzpicture}
			\caption{Illustration of the numerical fluxes used to update a intermediate cell.}
			\label{fig: Intermediate Cell Update}
		\end{figure}
		The same procedure for updating a normal cell is applied to an intermediate cell however, due to the fact that fluxes at a intermediate interface are calculated on a sub-cell basis (Figure \ref{fig: Intermediate Cell Update}), they must first be summated. For this case illustrated in Figure \ref{fig: Intermediate Cell Update} this is equal to:
		\begin{equation}
			\textbf{F}_{i-\frac{1}{2}} = \sum_{k=1}^{3} \textbf{F}_{i-\frac{1}{2},k}
		\end{equation}
		\subsubsection{Structure Cells}
		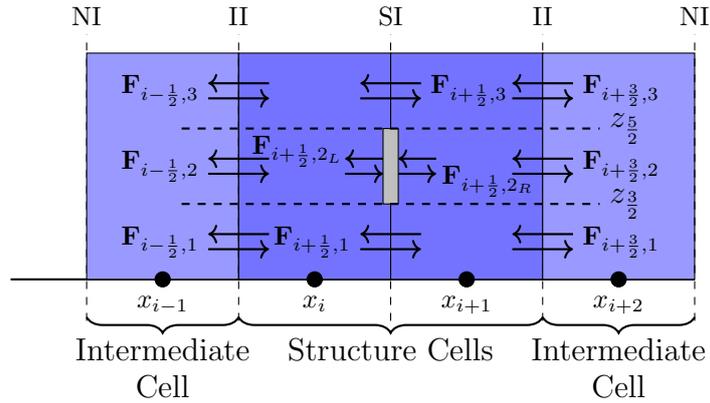
\begin{figure}[hbt!]
			\centering
			\begin{tikzpicture}
				\draw[thick] (,0) -- (10,0); 
				
				\draw[dashed] (2,-0.55) -- (2,3.25) node[above] {NI}; 
				\fill[fill=blue!40!white, draw=black] (2,0) rectangle (4,3); 
				\draw[dashed] (4,-0.55) -- (4,3.25) node[above] {II}; 
				\draw[dashed] (6,-0.55) -- (6,3.25) node[above] {SI}; 
				
				\fill[fill=blue!55!white, draw=black] (4,0) rectangle (6,3); 
				\fill[fill=blue!55!white, draw=black] (6,0) rectangle (8,3); 
				\fill[fill=black!25!white, draw=black] (5.9,1) rectangle (6.1,2); 
				
				\draw[dashed] (8,-0.55) -- (8,3.25) node[above] {II}; 
				\fill[fill=blue!40!white, draw=black] (8,0) rectangle (10,3); 
				\draw[dashed] (10,-0.55) -- (10,3.25) node[above] {NI}; 
				
				\draw[thick, dashed] (3.25, 1) -- (8.75, 1) node[right] {\large$z_\frac{3}{2}$};
				\draw[thick, dashed] (3.25, 2) -- (8.75, 2) node[right] {\large$z_\frac{5}{2}$};
				\fill[fill=black, draw=black] (5,0) circle (0.1cm);
				\node[below] at (5,-0.1) {$x_{i}$};
				\fill[fill=black, draw=black] (7,0) circle (0.1cm);
				\node[below] at (7,-0.1) {$x_{i+1}$};
				\draw [decorate, decoration={brace,amplitude=6pt,raise=0pt}, thick] (8,-0.5) -- (4,-0.5); 
				\node[below] at (6,-0.65) {\large Structure Cells};
				\fill[fill=black, draw=black] (3,0) circle (0.1cm);
				\node[below] at (3,-0.1) {$x_{i-1}$};
				\draw [decorate, decoration={brace,amplitude=6pt,raise=0pt}, thick] (4,-0.5) -- (2,-0.5); 
				\node[below] at (3,-0.65) {\large Intermediate};
				\node[below] at (3,-1.15) {\large Cell};
				\fill[fill=black, draw=black] (9,0) circle (0.1cm);
				\node[below] at (9,-0.1) {$x_{i+2}$};
				\draw [decorate, decoration={brace,amplitude=6pt,raise=0pt}, thick] (10,-0.5) -- (8,-0.5); 
				\node[below] at (9,-0.65) {\large Intermediate};
				\node[below] at (9,-1.15) {\large Cell};
				\draw[->, thick] (3.6,0.4) -- (4.4,0.4); 
				\draw[->, thick] (4.4,0.6) -- (3.6,0.6); 
				\node[left] at (3.6, 0.5) {$\textbf{F}_{i-\frac{1}{2},1}$};
				\draw[->, thick] (3.6,1.4) -- (4.4,1.4); 
				\draw[->, thick] (4.4,1.6) -- (3.6,1.6); 
				\node[left] at (3.6, 1.5) {$\textbf{F}_{i-\frac{1}{2},2}$};
				\draw[->, thick] (3.6,2.4) -- (4.4,2.4); 
				\draw[->, thick] (4.4,2.6) -- (3.6,2.6); 
				\node[left] at (3.6, 2.5) {$\textbf{F}_{i-\frac{1}{2},3}$};
				\draw[->, thick] (5.6,0.4) -- (6.4,0.4); 
				\draw[->, thick] (6.4,0.6) -- (5.6,0.6); 
				\node[left] at (5.6, 0.5) {$\textbf{F}_{i+\frac{1}{2},1}$};
				\node[left] at (5.5, 1.7) {$\textbf{F}_{i+\frac{1}{2},2_L}$};
				\draw[->, thick] (5.4,1.4) -- (5.9,1.4); 
				\draw[->, thick] (5.9,1.6) -- (5.4,1.6); 
				\node[right] at (6.55, 1.3) {$\textbf{F}_{i+\frac{1}{2},2_R}$};
				\draw[->, thick] (6.1,1.4) -- (6.6,1.4); 
				\draw[->, thick] (6.6,1.6) -- (6.1,1.6); 
				\node[right] at (6.4, 2.5) {$\textbf{F}_{i+\frac{1}{2},3}$};
				\draw[->, thick] (5.6,2.4) -- (6.4,2.4); 
				\draw[->, thick] (6.4,2.6) -- (5.6,2.6); 
				\draw[->, thick] (7.6,0.4) -- (8.4,0.4); 
				\draw[->, thick] (8.4,0.6) -- (7.6,0.6); 
				\node[right] at (8.4, 0.5) {$\textbf{F}_{i+\frac{3}{2},1}$};
				\draw[->, thick] (7.6,1.4) -- (8.4,1.4); 
				\draw[->, thick] (8.4,1.6) -- (7.6,1.6); 
				\node[right] at (8.4, 1.5) {$\textbf{F}_{i+\frac{3}{2},2}$};
				\draw[->, thick] (7.6,2.4) -- (8.4,2.4); 
				\draw[->, thick] (8.4,2.6) -- (7.6,2.6); 
				\node[right] at (8.4, 2.5) {$\textbf{F}_{i+\frac{3}{2},3}$};
			\end{tikzpicture}
			\caption{Illustration of the numerical fluxes used for updating the sub-cells of which a structure cells is comprised.}
			\label{fig: Structure Cell Update}
		\end{figure}
		Since structure cells are divided into sub-cells, it is necessary to update each individual sub-cell using the respective left and right fluxes as per:
		\begin{equation}\label{eq: Sub-Cell Update}
			\textbf{U}_{i,k}^{n+1} = \textbf{U}_{i,k}^{n} - \frac{\Delta t}{\Delta x}\left[\textbf{F}_{i+\frac{1}{2},k}-\textbf{F}_{i-\frac{1}{2},k}\right] + \Delta t \textbf{S}\left(\textbf{U}_{i,k}^n\right)
		\end{equation}
		Where $\textbf{U}_{i,k}^n$ represents the vector of conserved variables for the $k$th sub-cell contained within the $i$th cell at time level $n$. $\textbf{F}_{i-1/2,k}$ and $\textbf{F}_{i+1/2,k}$ represent the left and right fluxes for the $k$th layer of the $i$th cell. As for the normal cells, a well-balanced treatment of the topographic source terms may be achieved via the hydrostatic reconstruction method or via upwinding of the source terms. The remaining source terms may be treated using suitable explicit or implicit methods depending on the desired stability and constraint of the timestep. For strong stability and the convenience of advancing the solution at the timestep for the advection problem, a point implicit scheme is recommended for the friction source terms:
		\begin{equation}\label{eq: Point Implicit Friction Source Term}
			q^{n+1}_i = q^{n}_i + \Delta t \left(\frac{\left(\tau_{i,k+\frac{1}{2}}^{n+1}-\tau_{i,k-\frac{1}{2}}^{n+1}\right)}{1+\Delta t \left(\left(\frac{\partial \tau_{i,k+1/2}}{\partial q_{i,k}}\right)^{n} - \left(\frac{\partial \tau_{i,k-1/2}}{\partial q_{i,k}}\right)^{n} \right)} \right)
		\end{equation}
		At the sub-cell interfaces containing structures there are two numerical fluxes as illustrated in Figure \ref{fig: Structure Cell Update}, as a consequence of the two reflective boundaries implemented at each side of the structure. Since not all of the external forces are accounted for, these fluxes may be unequal, with the difference in the sum of the fluxes at the left face of the structure interface ($\textbf{F}^{(-)}$) and the right face of the structure interface ($\textbf{F}^{(-)}$) equal to the resultant hydrostatic pressure force exerted on the structure multiplied by the ratio of the timestep to the cell width ($\Delta t \backslash \Delta x (\textbf{F}^{(+)}-\textbf{F}^{(-)})$).
		
		Once the sub cells have been updated, their updated depth may exceed the maximum depth capacity for the layer and it is therefore necessary to re-define the layer properties of the structure cells in order to maintain alignment of the layers with the obstructions modelled at the interface. The process for redefining the layer properties is outlined in Algorithm \ref{Alg: Layer Redefinition}, for which an illustrative example is also provided via Figure \ref{fig: Layer Redefinition}.
		
		\begin{algorithm}
			\caption{Redefinition of the sub-cell properties based on the maximum depth capacity of the layers defined at a structure interface, post updating of the conserved variables. $\bar{h}$ and $\bar{q}$ represent the redefined depth and momentum. $j$ refers to the index of the redefined layers and $k$ refers to the index of the updated layer properties pre-redefinition. $n$ is the maximum number of layers defined at a structure interface.}\label{Alg: Layer Redefinition}
			\For{each structure cell}{
				$j \leftarrow 1$ \\ 
				$k \leftarrow 1$ \\
				$\bar{\textbf{h}}_j \leftarrow [0,...,0]$ \\
				$\bar{\textbf{q}}_j \leftarrow [0,...,0]$ \\
				\While{$\textrm{sum}(\bar{\textbf{h}}_j) < \textrm{sum}(\textbf{h}_k)$}{
					$h_{max} \leftarrow z_j - z_{j-1}$ \\
					\While{$\bar{h}_j < h_{max} \wedge k \leq n$}{
						$\bar{h}_j \leftarrow \bar{h}_j + h_k$ \\
						$\bar{q}_j \leftarrow \bar{q}_j + q_k$ \\
					$k \leftarrow k + 1$
					}
					$h_{excess} \leftarrow max(\bar{h}_j - h_{max}, 0)$ \\ 
					$\bar{h}_j \leftarrow \bar{h}_j -h_{excess}$ \\
					$\bar{q}_j \leftarrow \bar{q}_j - h_{excess}u_{k-1}$ \\
					$\bar{h}_{j+1} \leftarrow h_{excess}$ \\
					$\bar{q}_{j+1} \leftarrow h_{excess}u_{k-1}$ \\
					$j \leftarrow j + 1$
				}
			}
		\end{algorithm}
		
		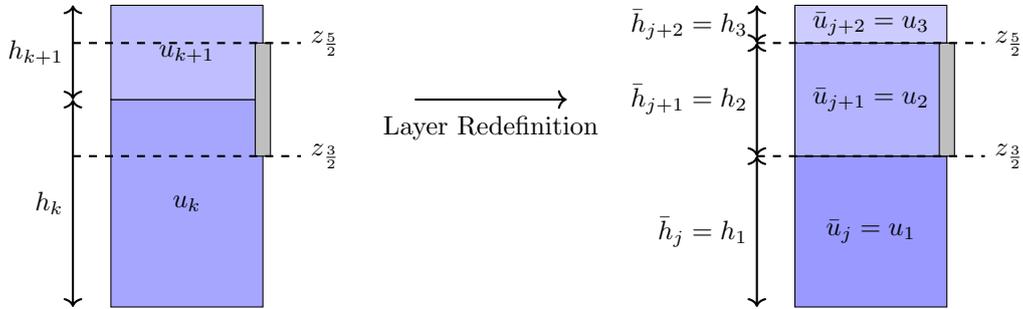
\begin{figure}[hbt!]
			\centering
			\begin{tikzpicture}
			\fill[fill=blue!35!white, draw=black] (0,0) rectangle (2,2.75);
			\fill[fill=blue!25!white, draw=black] (0,2.75) rectangle (2,4);
			\fill[fill=black!25!white, draw=black] (1.9,2) rectangle (2.1,3.5);
			\draw[thick, dashed] (-0.5,2) -- (2.5,2) node[right] {$z_\frac{3}{2}$};
			\draw[thick, dashed] (-0.5,3.5) -- (2.5,3.5) node[right] {$z_\frac{5}{2}$};
			
			\draw[thick,<->] (-0.5,0) -- (-0.5,2.75);
			\draw[thick,<->] (-0.5,2.75) -- (-0.5,4);
			\node[left] at (-0.5,1.375) {$h_{k}$};
			\node[left] at (-0.5,3.375) {$h_{k+1}$};
			\node[] at (1,1.375) {$u_{k}$};
			\node[] at (1,3.35) {$u_{k+1}$};
			
			\draw[thick, ->] (4,2.75) -- (6,2.75);
			\node[below] at (5,2.65) {Layer Redefinition};
			
			\fill[fill=blue!40!white, draw=black] (9,0) rectangle (11,2);
			\fill[fill=blue!30!white, draw=black] (9,2) rectangle (11,3.5);
			\fill[fill=blue!20!white, draw=black] (9,3.5) rectangle (11,4);
			\fill[fill=black!25!white, draw=black] (10.9,2) rectangle (11.1,3.5);
			\draw[thick, dashed] (8.5,2) -- (11.5,2) node[right] {$z_\frac{3}{2}$};
			\draw[thick, dashed] (8.5,3.5) -- (11.5,3.5) node[right] {$z_\frac{5}{2}$};
			
			\draw[thick,<->] (8.5,0) -- (8.5,2);
			\draw[thick,<->] (8.5,2) -- (8.5,3.5);
			\draw[thick,<->] (8.5,3.5) -- (8.5,4);
			\node[left] at (8.5,1) {$\bar{h}_{j}=h_1$};
			\node[left] at (8.5,2.75) {$\bar{h}_{j+1}=h_2$};
			\node[left] at (8.5,3.75) {$\bar{h}_{j+2}=h_3$};
			\node[] at (10,1) {$\bar{u}_{j}=u_1$};
			\node[] at (10,2.75) {$\bar{u}_{j+1}=u_2$};
			\node[] at (10,3.75) {$\bar{u}_{j+2}=u_3$};
			\end{tikzpicture}
			\caption{Illustration of the layer redefinition process post updating of the conserved variables. The redefinition process is required to re-align the updated properties of the sub-cells with the respective boundary conditions implemented at the structure interface.}
			\label{fig: Layer Redefinition}
		\end{figure} \newpage
		\section{Model Validation}
		\begin{figure}[hbt!]	
			\begin{center}
				\includegraphics[width=.75\linewidth]{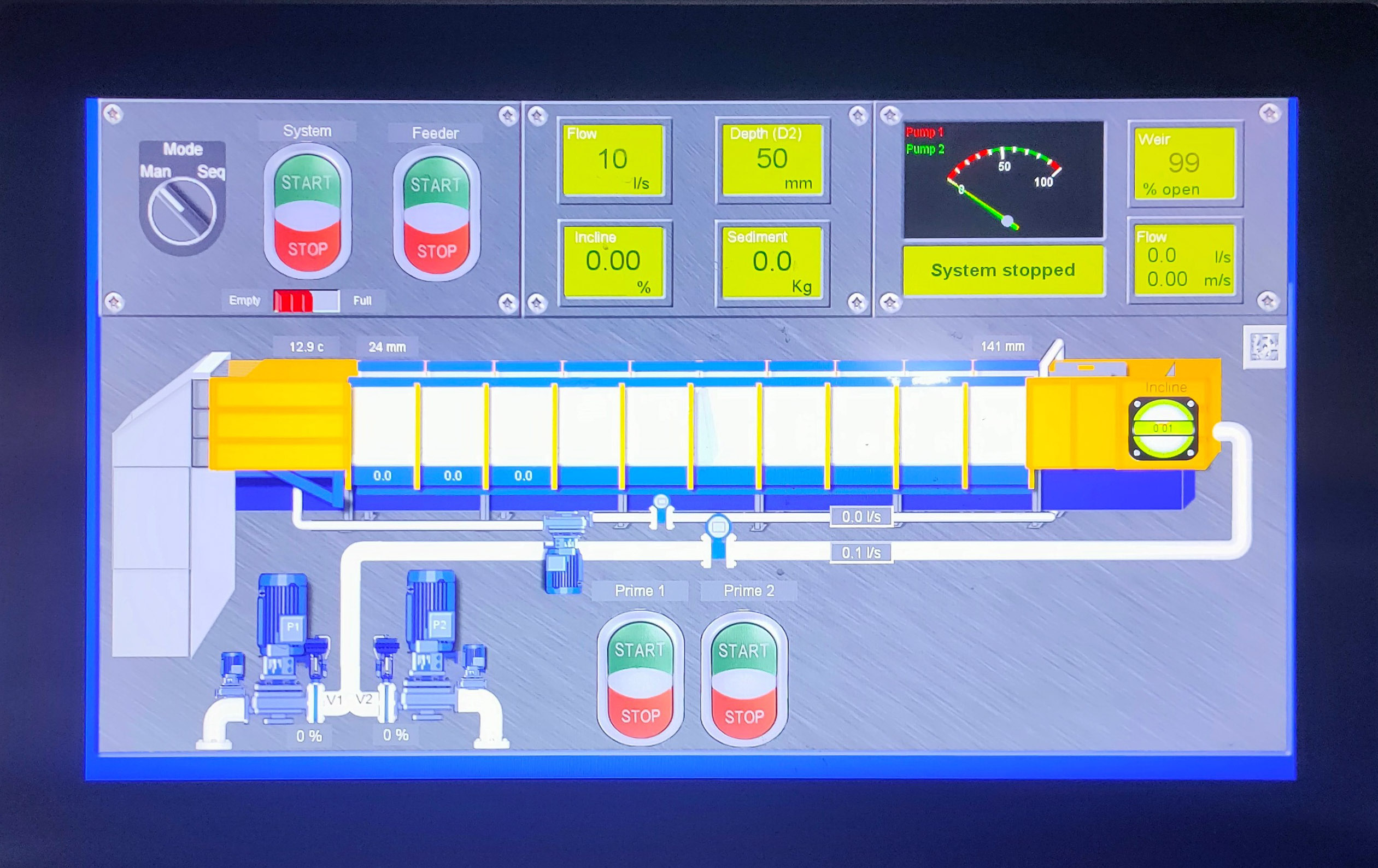}
			\end{center}
			\caption{Integrated control panel for the S100 Research Flume, including a schematic of the flume. Two pumps, which draw water from a recirculating sump, supply water to the flume via a pipe connected to the upstream (right) end. At the left end of the flume, the water exits the flume via a sloped free outfall into the the sump.}
		\label{Fig: Flume Control Panel}
		\end{figure}
		Previously published validation data \cite{RN100}, collected from experiments conducted in Newcastle University's Armfield S100 Research Flume, is used to validate the accuracy of the proposed Riemann solver. The S100 Research Flume is a $12.5$m long, $1$m wide, $0.8$m deep flume capable of producing flow rates up to $400$ls$^{-1}$. Using the control panel, shown in Figure \ref{Fig: Flume Control Panel}, the user can select a desired flow rate which is then produced by the two pumps which draw water from the sump. The flow rate is maintained and corrected via a  proportional-integral-derivative control loop, which uses a electromagnetic flow meter (Euromag Model MUT2200EL) to ensure that flow rate within the inflow pipe matches the desired flow rate. According to Euromag technical sheet \cite{RN69}, each sensor is calibrated on a hydraulic test rig equipped with an ISO17025 traceable weighing system, which ensures that the accuracy of the sensor is equal to $0.2\%\pm2$mms$^{-1}$ with a repeatability of approximately $0.1\%$. A summary of the maximum permissible error limits for the instrument, provided by the manufacturer, is presented in Table \ref{Table: EuroMag Accuracy}.
		\begin{table}[ht]
			\begin{tabular}{ |c|c|c|c| }
				\hline
				\multicolumn{4}{|c|}{Maximum Permissible Error limits for Euromag Model MUT2200EL DN 350 PN 10 EN 1092-1} \\
				\hline
				Flow Rate & $q_1 = 12.800m^3h^{-1}$ & $q_2 = 20.480m^3h^{-1}$ & $q_3 = 360.000m^3h^{-1}$ \\
				\hline
				Instrument Error & $\pm \ 4.99\%$ & $\pm \ 2.00\%$ & $\pm \ 0.49\%$ \\
				\hline
			\end{tabular}
			\centering
			\caption{Maximum permissible error limits for the electromagnetic flow meter for a range of flow rates within the inflow pipe (adapted from \cite{RN69} p.4).}
			\label{Table: EuroMag Accuracy}
		\end{table} \\
		The validation experiments consisted of running the flume at a range of flow rates, with a range of different barrier geometries placed within the flume cross-section, at a distance of $5$m downstream. The flume tilt was set to $0\%$ for all validation experiments in order to eliminate any potential numerical errors introduced as a result of topographic source terms. Once steady state flow conditions were achieved for each experiment, depth measurements were obtained using vernier point gauges. The full validation dataset is available as supplementary material from the referenced publication. 
		\subsection{Numerical Setup}\label{section: Numerical Setup}
		All numerical simulations were conducted on a $12.5$m 1D spatial domain, discretised into a structured grid comprised of $0.1$m cells ($\Delta x = 0.1$m). In order to ensure satisfaction of the Courant-Friedrichs-Lewy condition, a Courant number of $C = (0.95\Delta x)/(S^n_{max})$ was used to determine a stable timestep, where $S_{max}^n$ is the maximum absolute wave speed at time level $n$. Since the bed slope is set to $0\%$ this has the intended effect of simplifying the source terms, only requiring the friction source term to be resolved, facilitating clearer analysis of the accuracy of the Riemann solver. The friction source terms for normal and intermediate cells are resolved using (\ref{eq: Friction Source Term}). The friction source terms for the structure cells are resolved using (\ref{eq: Point Implicit Friction Source Term}). A Manning's n equal to $0.012$ and a kinematic viscosity of $1.0034\times10^{-6}$m$^2$s$^{-1}$ is assumed for all numerical simulations. 

		The upstream and downstream boundary conditions are both implemented using exterior ghost cells. In order to replicate the constant inflow produced by the S100 flume, an inflow boundary condition is defined at the upstream end utilising relationships derived from the Riemann invariants across a rarefaction wave. At the downstream boundary a critical depth boundary condition is imposed. Full details for the implementation of the boundary conditions are presented in \cite{RN100}.
		\section{Results}
		The following validation test cases can be categorised into three primary flow configurations:
		\begin{itemize}
			\item Flow under a barrier.
			\item Flow under a barrier, producing a downstream stationary hydraulic jump.
			\item Flow over and under a barrier.
		\end{itemize}
		Through comparisons between the experimental and numerical data for the six presented validation test cases, the suitability and accuracy of the proposed solver is demonstrated. 
		\subsubsection{Flow Under a Barrier}
		For test case one and test case two, the solver produced accurate predictions for the upstream and downstream depth, capturing the interaction of the flow with the obstruction. In both test cases there is a slight overestimation of the upstream depth which equated to an error of $0.7-8.3\%$ for test case one and $0.1-12.6\%$ for test case two. In contrast, the downstream depth was slightly overestimated in both test cases, with a greater error for test case two due to the numerical prediction of a hydraulic jump at approximately $x=10m$ downstream. This is potentially a consequence of greater uncertainty in the measurement of the downstream depth, due to the presence of turbulent and unsteady flow at the outfall, which contributed to difficulty implementing the correct downstream boundary condition. Moreover, the location of a stationary hydraulic jump was determined to be extremely sensitive to small deviations in the flow during the execution of the lab experiments.
		
		The velocity upstream of the barrier is predicted accurately for both test cases with errors in the region of $7-11\%$. The numerical estimation of the velocity at the upstream face of the barrier has a larger error however, this is a localised error, constrained only to the structure cell immediately upstream of the barrier. Since the discharge predictions are accurate otherwise, the overestimation of the downstream depth corresponds to an underestimation of the downstream velocity equating to an error of $2.4-18.8\%$ for test case one and $0.8-20.6\%$ for test case two. With the larger errors for test case two arising as a result of the incorrect prediction of the hydraulic jump.
		\begin{figure}[hbt!]\centering
			\begin{subfigure}{.85\linewidth}
				\centering
				\includegraphics[width=.85\linewidth]{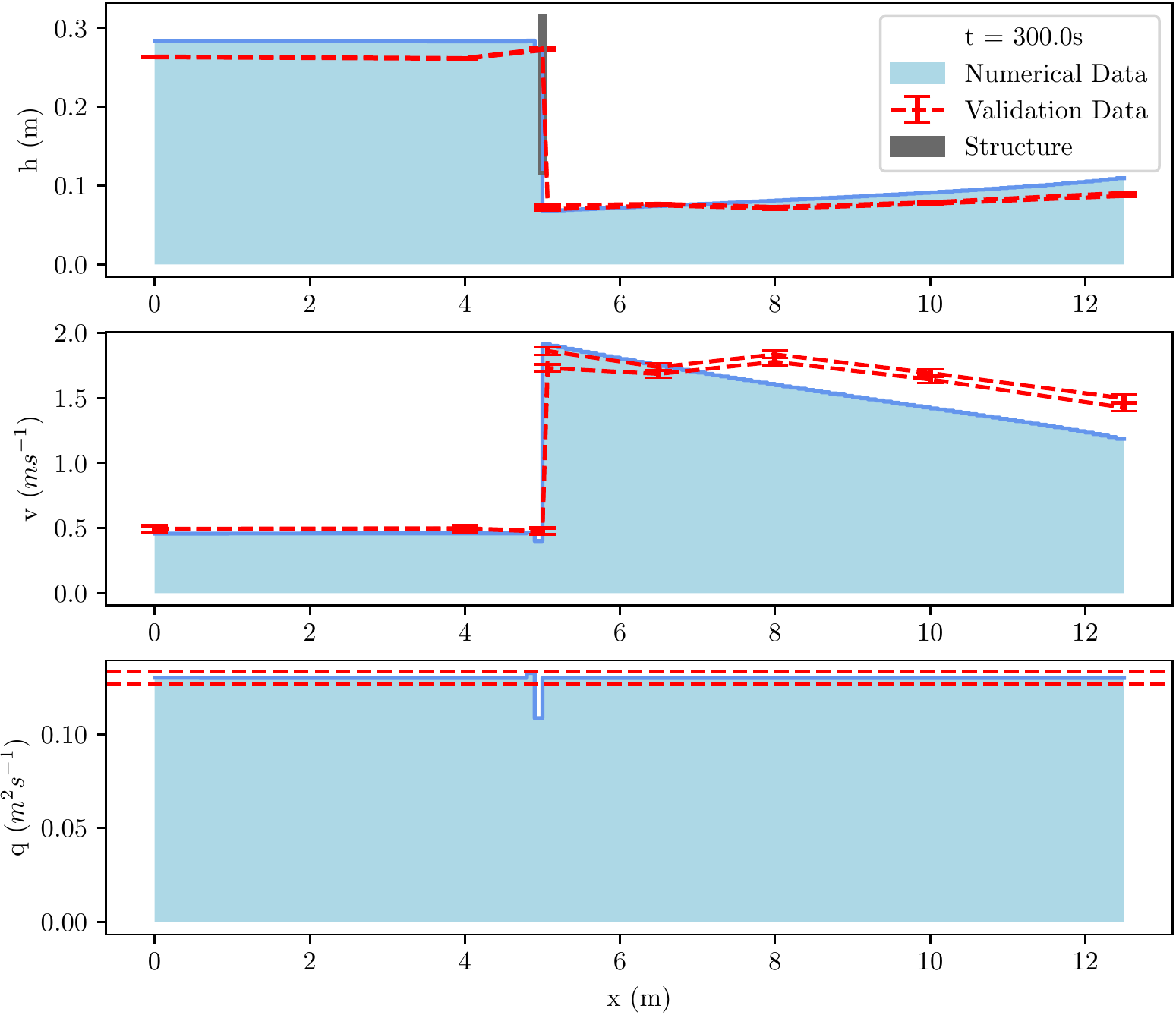}
				\caption{}\label{Fig: Test Case 1 Plot}
			\end{subfigure}
			\begin{subfigure}{.85\linewidth}
				\centering
				\includegraphics[width=0.85\textwidth]{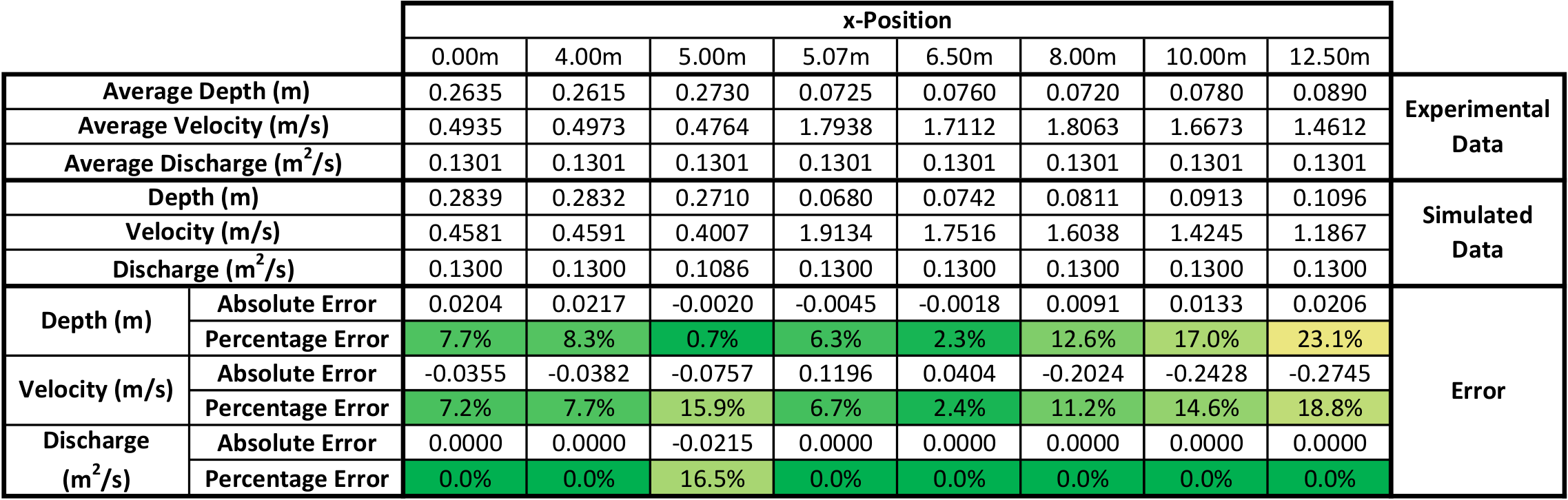}
				\caption{}\label{Fig: Test Case 1 Error Table}
			\end{subfigure}
			\caption{Comparison between numerical and experimental results for test case $1$. Details of the numerical setup can be found in Section \ref{section: Numerical Setup}.}
			\label{Fig: Test Case 1}
		\end{figure} \clearpage
	
		\begin{figure}[hbt!]\centering
			\begin{subfigure}{.85\linewidth}
				\centering
				\includegraphics[width=.85\linewidth]{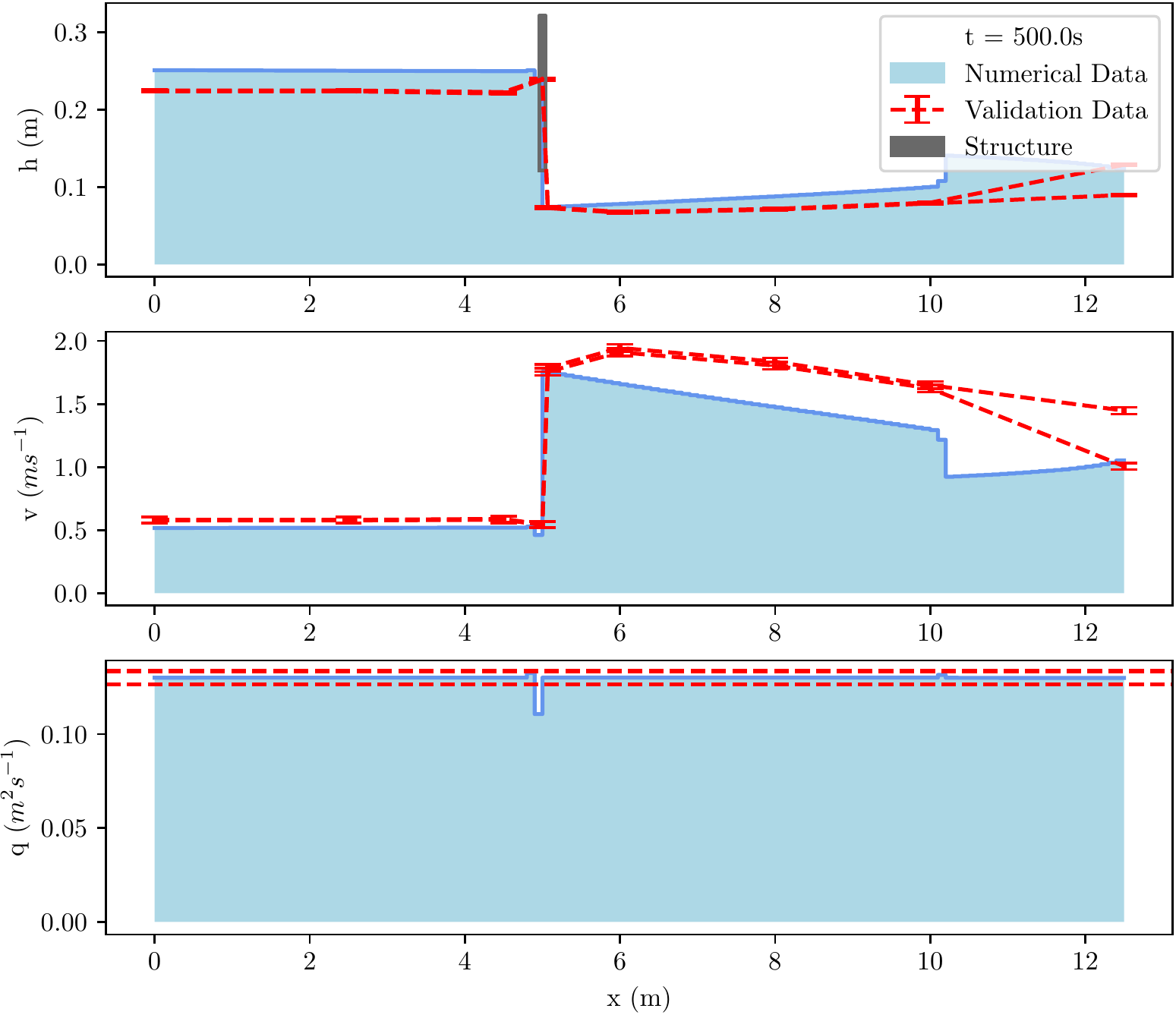}
				\caption{}\label{Fig: Test Case 2 Plot}
			\end{subfigure}
			\begin{subfigure}{.85\linewidth}
				\centering
				\includegraphics[width=0.85\textwidth]{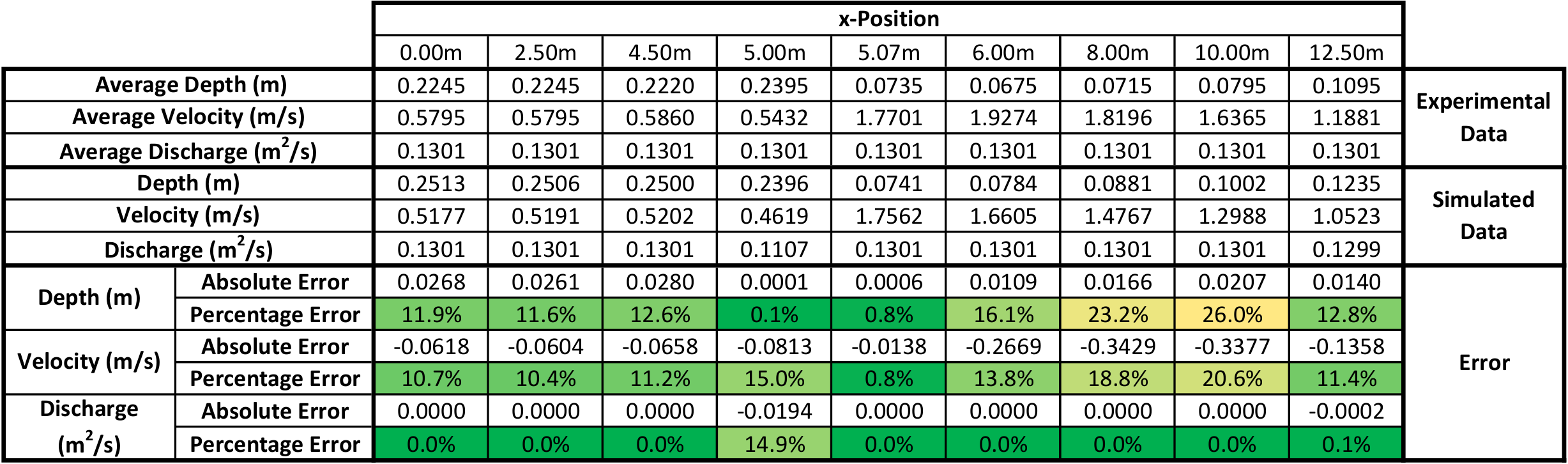}
				\caption{}\label{Fig: Test Case 2 Error Table}
			\end{subfigure}
			\caption{Comparison between numerical and experimental results for test case $2$. Details of the numerical setup can be found in Section \ref{section: Numerical Setup}.}
			\label{Fig: Test Case 2}
		\end{figure} \clearpage
		\subsubsection{Stationary Hydraulic Jump}
		Test case five and test case six showcase the capacity of the solver to accurately resolve stationary hydraulic jumps. The two presented test cases use the same barrier configurations with different flow rates, resulting in the formation of different stationary hydraulic jumps for each scenario. In both cases, the numerical results correctly predicted the formation of a stationary hydraulic jump downstream of the barrier. For test case six, the position and height of the jump was accurately captured. For test case five, the height of the jump was accurately captured, however, the formation of the jump was premature occurring at approximately $x=0.5$m upstream of the actual location. The robust wave estimation algorithm (Algorithm \ref{Alg: Wavespeeds}) was determined to be crucial for accurately capturing and maintaining the stationary hydraulic jumps for the relevant numerical simulations.
		
		In both cases the numerical results predict jumps with a zero length roller, characterised by a sharp discontinuity in the depth of flow at the toe of the jump, which is a feature of the classical shallow water equations; since there is no internal energy transcribed within the classical shallow water equations, energy loss through a shock discontinuity is instead captured via Rankine-Hugoniot relations arising from the conservation of mass and momentum \cite{RN307}. This is insufficient to capture the complex behaviour which occurs within the transition region of turbulent hydraulic jumps with a Froude number of greater than $1.5$. Methods to overcome the shortcomings of the classical shallow water equations, such as the work of Richard and Gavrilyuk \cite{RN306}, are not appropriate nor necessary for the desired application of flood risk modelling. 
		
		More generally, the predictions of the upstream and downstream depth and velocity proved to be accurate for both test cases, outside of the early prediction of the hydraulic jump for test case five. For test case five, there was a slight over estimation of the upstream depth corresponding to an error in the region of $1.4-12.3\%$. Ignoring the region of the domain occupied by the hydraulic jump ($6-7$m), downstream depth predictions were also found to be accurate with errors in the region of $0.7-13.2\%$. For test case six, the accuracy of the predictions starts to degrade towards the downstream boundary suggesting that the boundary condition may not be optimal. However, despite the increasing errors towards the downstream boundary, the solver still contributed to accurate results overall with depth errors from $0.2-19.2\%$ and velocity errors from $0.2-23.6\%$ for the data points between $x=0-8$m, with errors increasing to $27.3\%$ and $36.7\%$ respectively at the boundary.
		\begin{figure}[hbt!]\centering
			\begin{subfigure}{.85\linewidth}
				\centering
				\includegraphics[width=.85\linewidth]{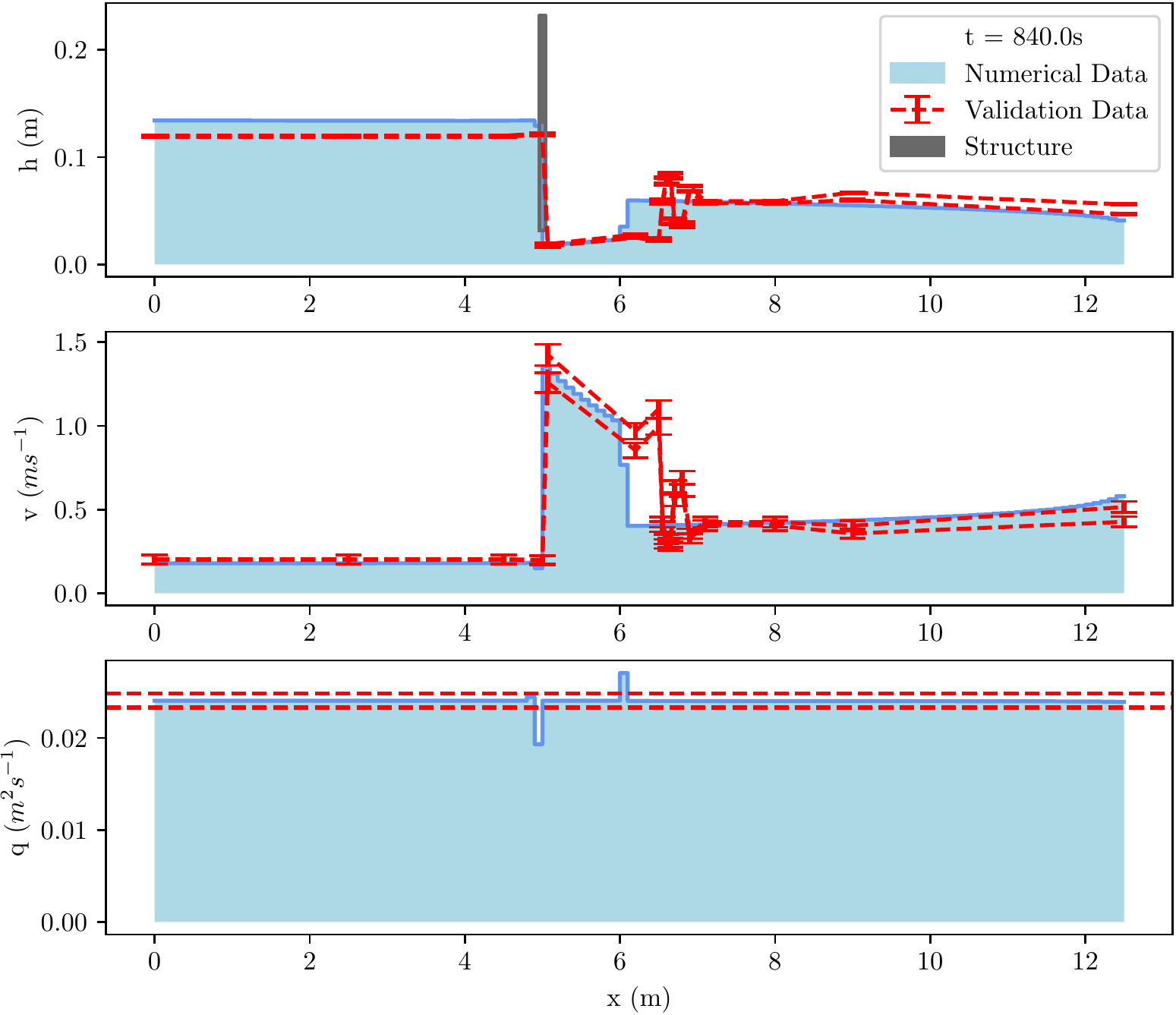}
				\caption{}\label{Fig: Test Case 5 Plot}
			\end{subfigure}
			\begin{subfigure}{.85\linewidth}
				\centering
				\includegraphics[width=0.85\textwidth]{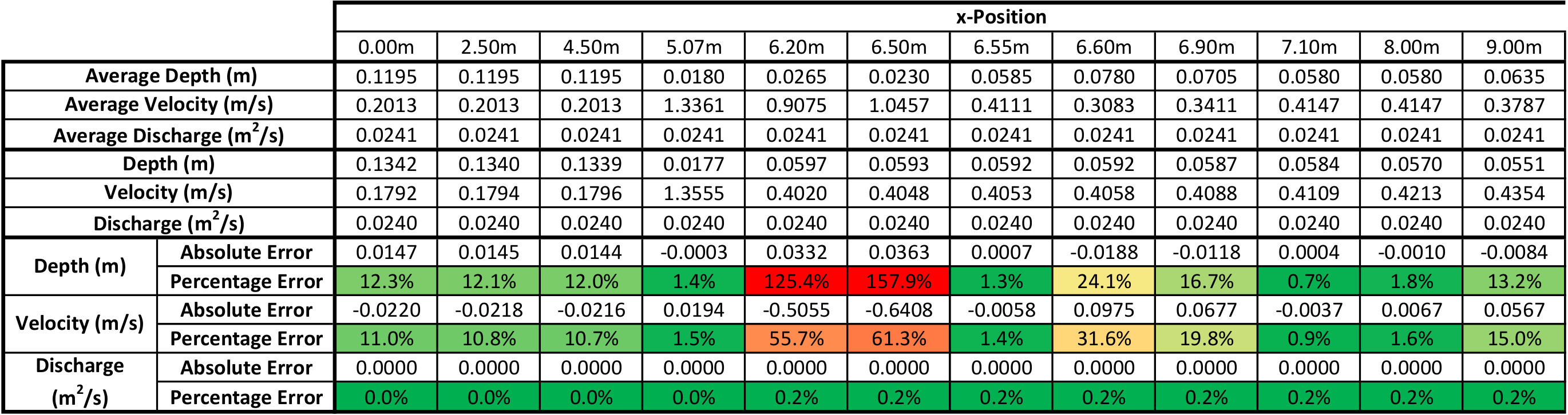}
				\caption{}\label{Fig: Test Case 5 Error Table}
			\end{subfigure}
			\caption{Comparison between numerical and experimental results for test case $5$. Details of the numerical setup can be found in Section \ref{section: Numerical Setup}.}
			\label{Fig: Test Case 5}
		\end{figure} \clearpage
		\begin{figure}[hbt!]\centering
			\begin{subfigure}{.85\linewidth}
				\centering
				\includegraphics[width=.85\linewidth]{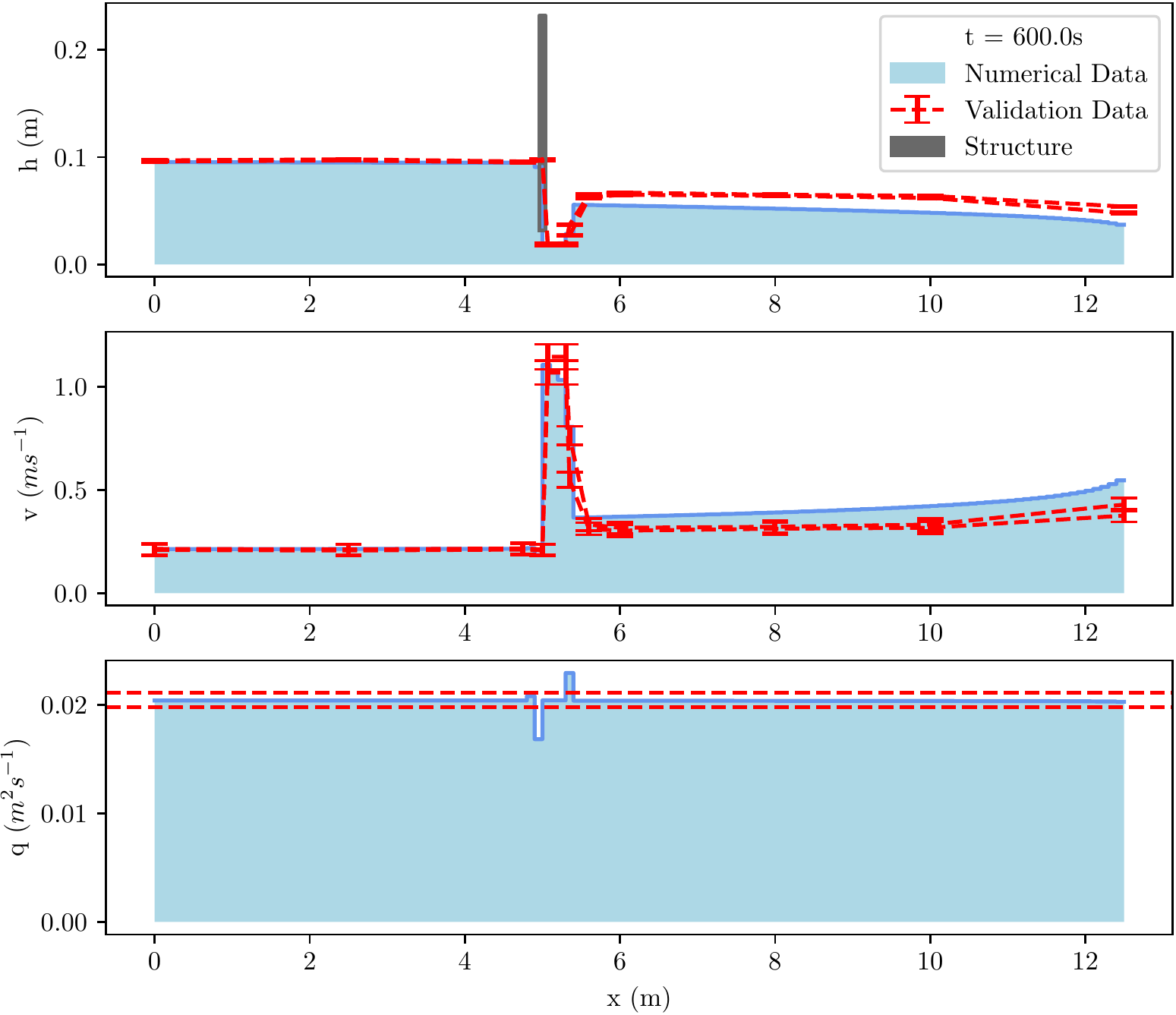}
				\caption{}\label{Fig: Test Case 6 Plot}
			\end{subfigure}
			\begin{subfigure}{.85\linewidth}
				\centering
				\includegraphics[width=0.85\textwidth]{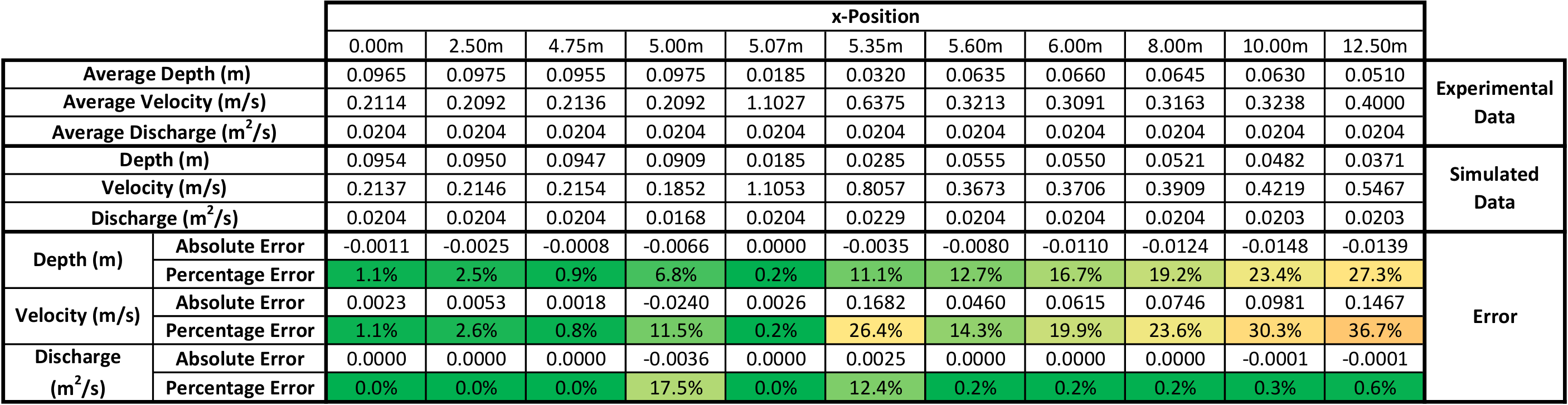}
				\caption{}\label{Fig: Test Case 6 Error Table}
			\end{subfigure}
			\caption{Comparison between numerical and experimental results for test case $6$. Details of the numerical setup can be found in Section \ref{section: Numerical Setup}.}
			\label{Fig: Test Case 6}
		\end{figure} \clearpage
		\subsubsection{Flow Over and Under a Barrier}
		For test case eight and test case nine, depth predictions proved to be accurate with errors increasing towards the downstream boundary in both cases. For test case eight, upstream depth predictions were extremely accurate ($0.1-3.3\%$). The upstream depth was overestimated for test case nine but remained accurate with errors in the region of $5.9-10.5\%$. For both test cases, there was a slight overestimation of the downstream depth with errors in the range of $6.5-24.7\%$. Figure \ref{Fig: Test Case 8} demonstrates that the water was observed as vertically flowing over the barrier for test case eight which cannot be captured by the numerical model, due to the nature of the fundamental equations and the structure of the finite volume scheme. Although this behaviour isn't captured by the model, the overall results remain accurate and the general behaviour is well captured. Certainly, for applications concerning flood risk modelling, the key quantities are the upstream and downstream depths which are observed to be consistent with the validation data.
		
		The velocity predictions are similarly accurate with errors in the region of $1.2-19.8\%$ for both of the presented test cases. As for the previous test cases, there is also a local error in the prediction of the discharge in the cells proceeding the barrier with discharge predictions otherwise proving accurate. 
		\begin{figure}[hbt!]\centering
			\begin{subfigure}{.85\linewidth}
				\centering
				\includegraphics[width=.85\linewidth]{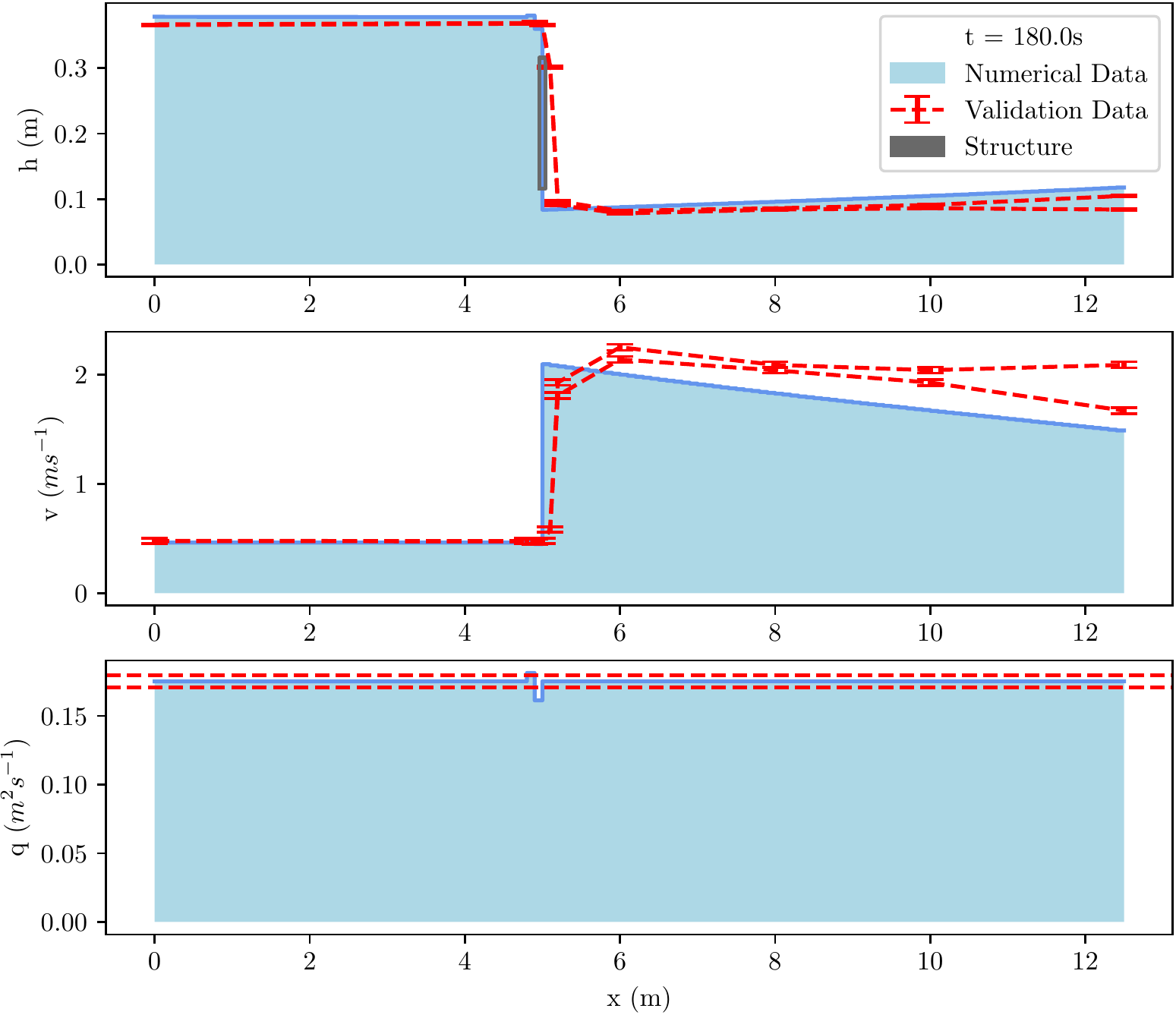}
				\caption{}\label{Fig: Test Case 8 Plot}
			\end{subfigure}
			\begin{subfigure}{.85\linewidth}
				\centering
				\includegraphics[width=0.85\textwidth]{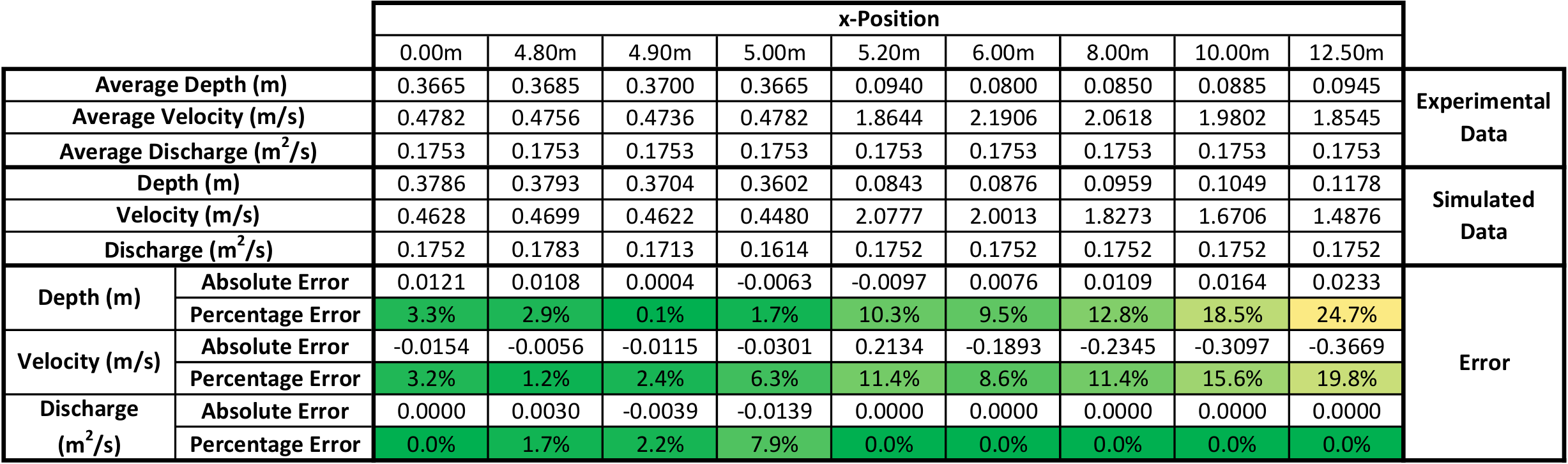}
				\caption{}\label{Fig: Test Case 8 Error Table}
			\end{subfigure}
			\caption{Comparison between numerical and experimental results for test case $8$. Details of the numerical setup can be found in Section \ref{section: Numerical Setup}.}
			\label{Fig: Test Case 8}
		\end{figure} \clearpage
		\begin{figure}[hbt!]\centering
			\begin{subfigure}{.85\linewidth}
				\centering
				\includegraphics[width=.85\linewidth]{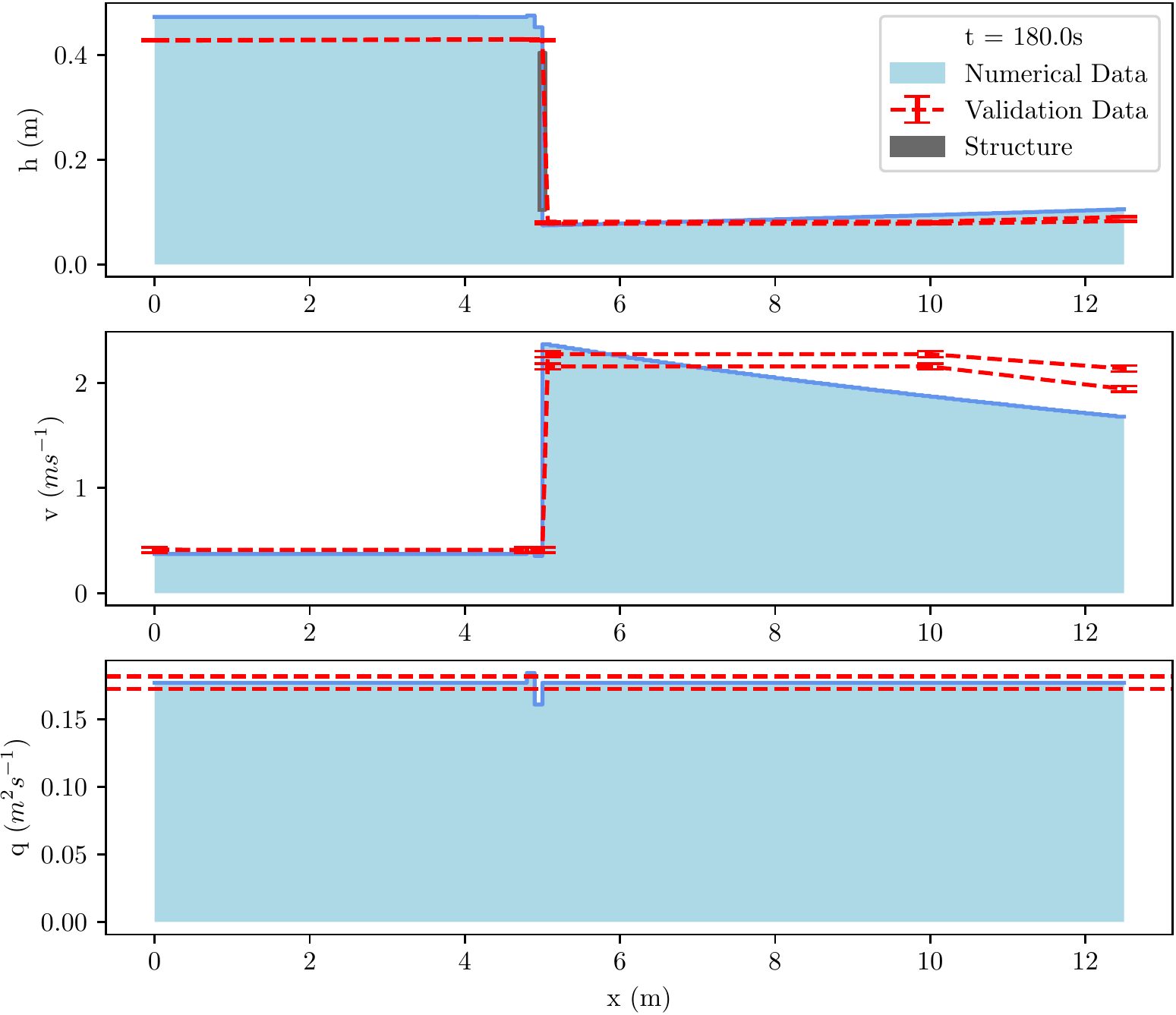}
				\caption{}\label{Fig: Test Case 9 Plot}
			\end{subfigure}
			\begin{subfigure}{.85\linewidth}
				\centering
				\includegraphics[width=0.85\textwidth]{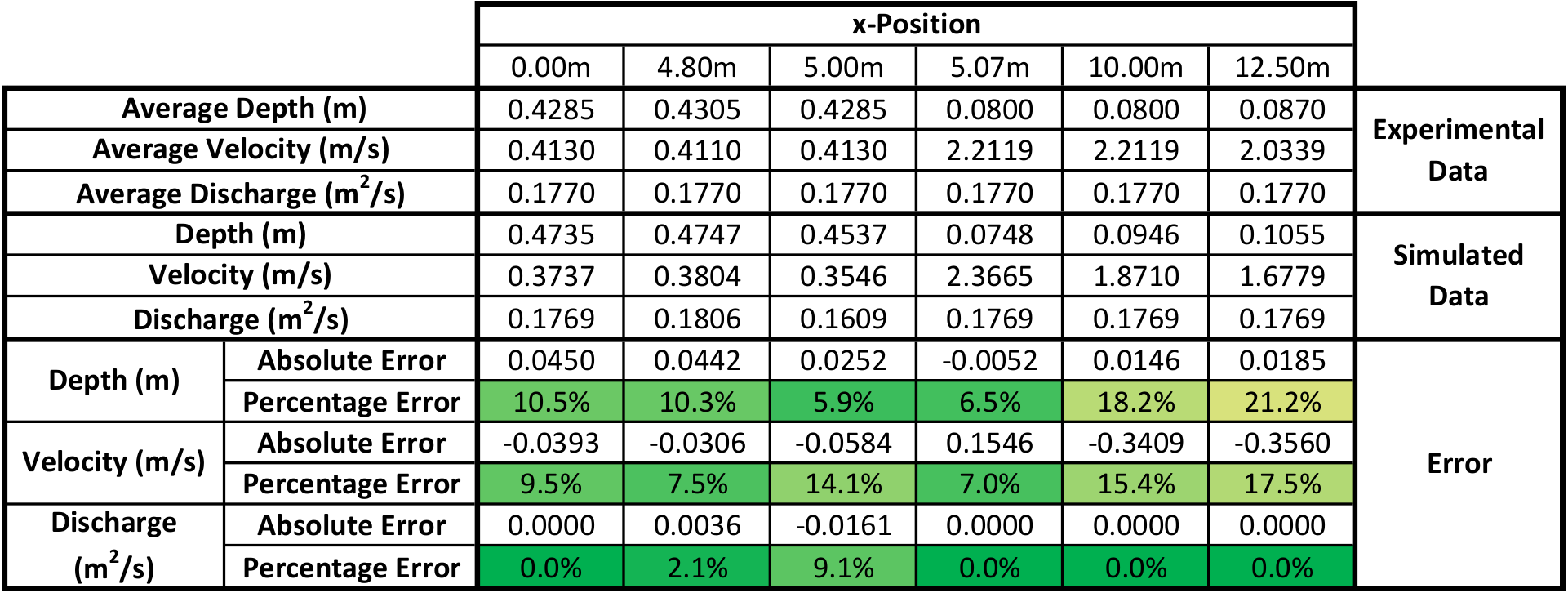}
				\caption{}\label{Fig: Test Case 9 Error Table}
			\end{subfigure}
			\caption{Comparison between numerical and experimental results for test case $9$. Details of the numerical setup can be found in Section \ref{section: Numerical Setup}.}
			\label{Fig: Test Case 9}
		\end{figure} \clearpage
		\subsubsection{Mesh Convergence Analysis}
		\begin{table}[hbt!]
			\begin{center}
				\includegraphics[width=.95\linewidth]{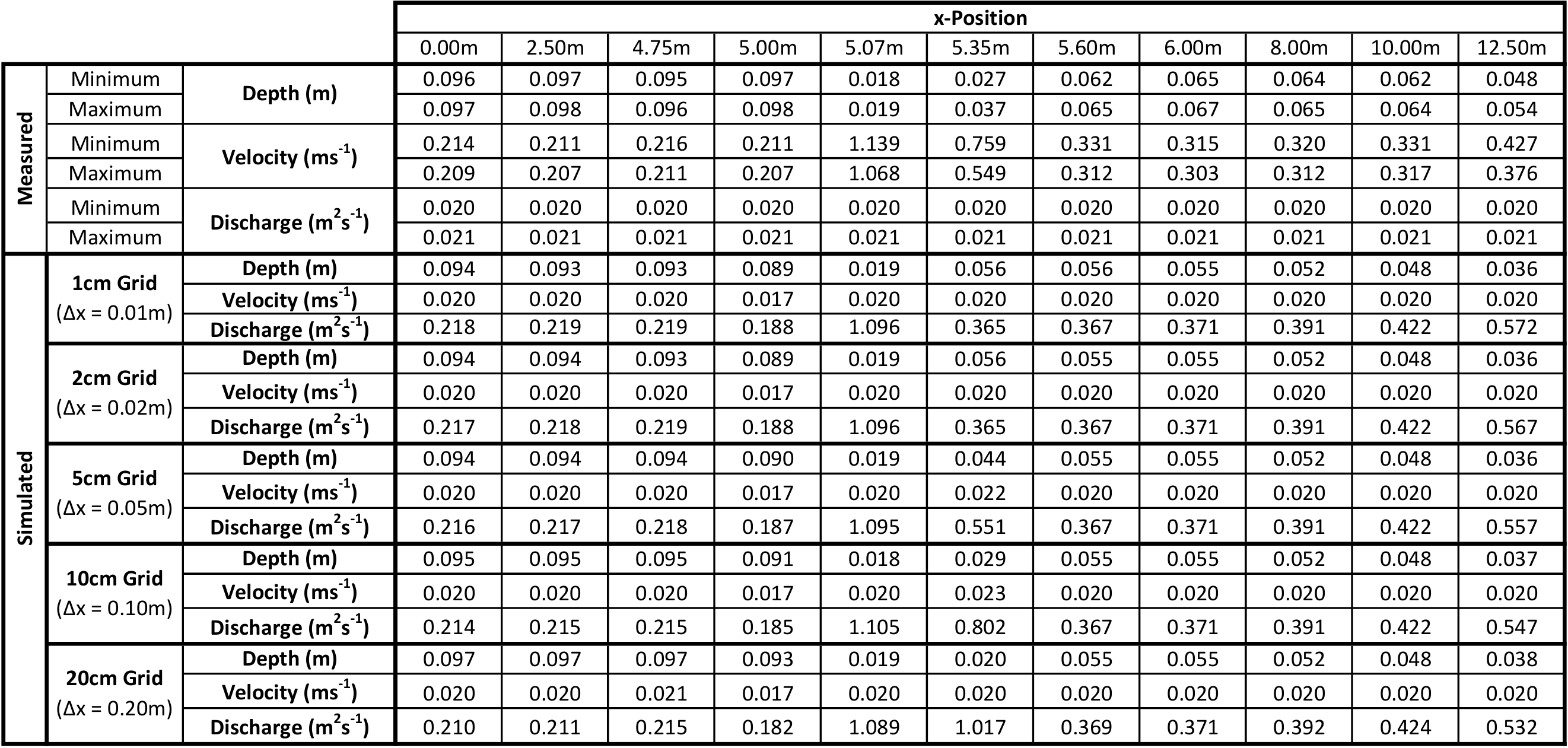}
			\end{center}
			\caption{Comparison between the experimental data and simulated data for Test Case 6, for a range of mesh resolutions.}
			\label{Table: Mesh Convergence}
		\end{table}
		The data in Table \ref{Table: Mesh Convergence} demonstrates negligible differences in the results for the tested mesh resolutions which range between $1-20$cm ($\Delta x = 1-20$cm). The relevant plots illustrating the results can be found in the Appendices. The primary difference between the meshes is in the sharpness of the depth discontinuity at the toe of the stationary hydraulic jump, which becomes steeper as the mesh is refined.
		\section{Comparison}
		In order to demonstrate the comparative value of the solver presented within this paper, designated as solver 2, a comparison is presented with the solver presented in \cite{RN100}, designated as solver 1. A comparison of the solvers for test case one, shown in Figure \ref{Fig: Test Case 1 Comparison}, demonstrates accurate results for both solvers. Solver 2 has a marked increase in accuracy for the depth upstream of the barrier, whereas, the depth and velocity downstream of the barrier is captured slightly more accurately by Solver 1. Similarly for test case 8, shown in Figure \ref{Fig: Test Case 8 Comparison}, there is a improvement in the prediction of the upstream depth for Solver 2, with comparatively accurate results for both solvers downstream of the barrier. The benefits of Solver 2 are however, showcased best via Figure \ref{Fig: Test Case 5 Comparison} and Figure \ref{Fig: Test Case 6 Comparison}. Whilst Solver 1 is able to broadly capture the upstream and downstream depths, which is of primary concern for flood risk management applications, Solver 2 captures the upstream and downstream depths more accurately including the formation of a stationary hydraulic jump. The difference in the accuracy of the velocity predictions downstream of the barrier is stark and demonstrates that the superiority of the solution procedure utilised by Solver 2 with regards to capturing the horizontal velocity profile in the vertical plane at the structure interface. 
		
		Although it is clear that Solver 2 is capable of producing superior results, there is a clear increase in complexity and computational burden in comparison with Solver 1. However, since the implementation is local, on a sufficiently large domain the difference in computational efficiency of the two solvers is unlikely to be a limiting factor since structure cells are likely to comprise a very small percentage of all cells within the computational domain. As such, the primary grounds for the use of Solver 1 over Solver 2 should be limited to scenarios in which the simplicity of implementation is important and for use cases which predominantly involve supercritical downstream flow regimes. Otherwise, Solver 2 proves to be the optimal solution. Furthermore, the capacity for Solver 2 to accurately resolve stationary hydraulic jumps and more accurately capture the velocity at the barrier presents further opportunities such as the modelling of the transport of water soluble contaminants. Passive scalars, such as water soluble contaminants, are assumed to be passively advected with the fluid and via the reintroduction of the contact discontinuity wave, via switching from a HLL to a HLLC (Harten-Lax-van Leer contact) approximate Riemann solver \cite{RN309}, their transport can be modelled. Since this process is highly dependent on the accurate determination of the velocity, this is only possible for Solver 2. This has important applications in terms of modelling water quality, especially since the combination of flows around obstacles and species equations is seldom explored and is therefore to be the subject of further work.
		
		\begin{figure}[hbt!]\centering
			\begin{subfigure}{.85\linewidth}
				\centering
				\includegraphics[width=.85\linewidth]{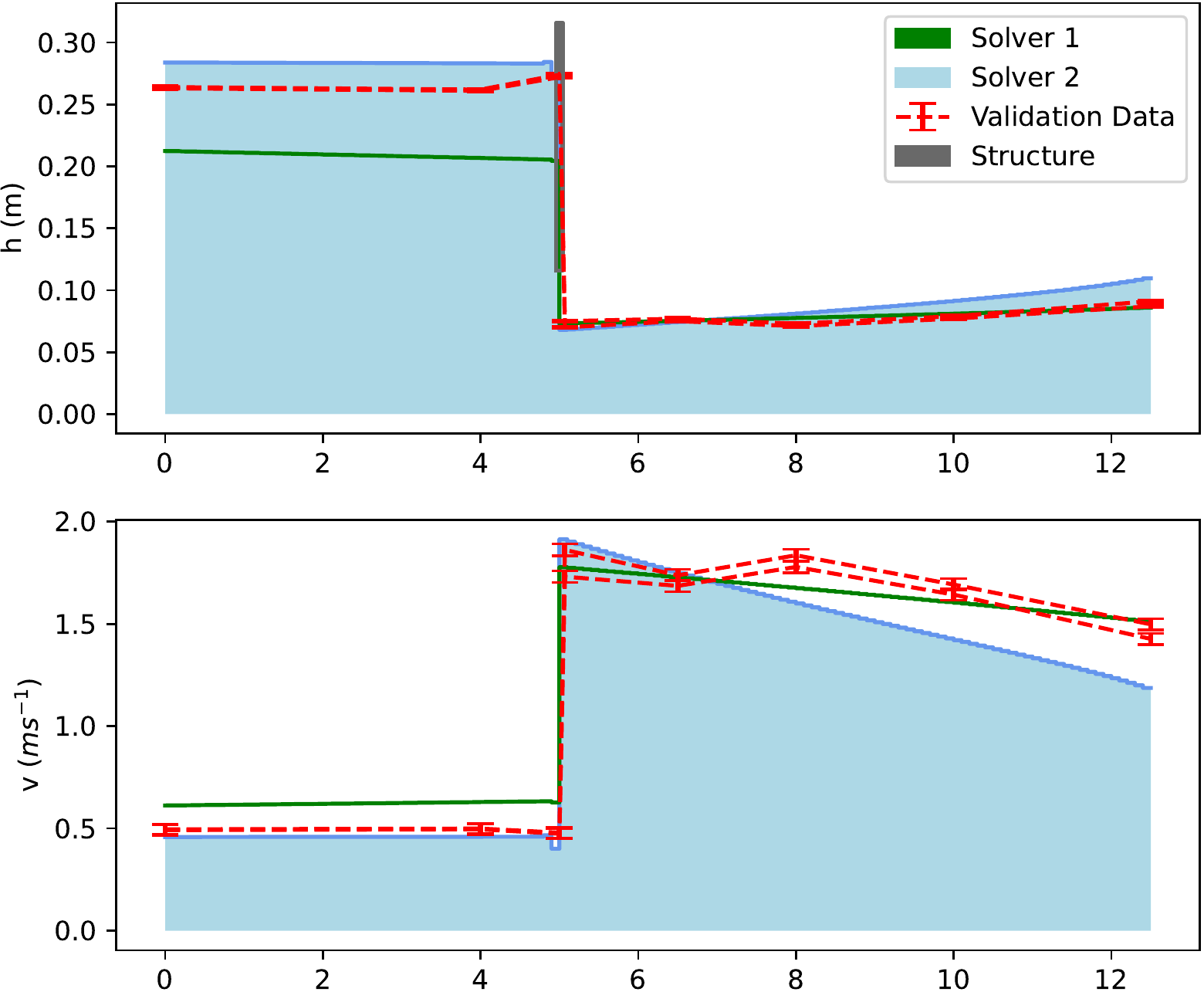}
				\caption{}\label{Fig: Test Case 1 Comparison Plot}
			\end{subfigure}
			\begin{subfigure}{.85\linewidth}
				\centering
				\includegraphics[width=0.85\textwidth]{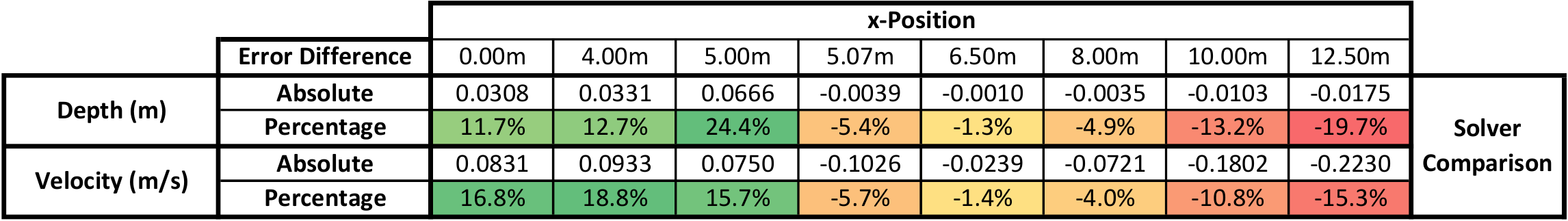}
				\caption{}\label{Fig: Test Case 1 Error Comparison}
			\end{subfigure}
			\caption{Comparison between solver 1 and solver 2 with respect to the experimental results for test case $1$. Details for solver 1 can be found in \cite{RN100}. Details for the numerical setup can be found in Section \ref{section: Numerical Setup}.}
			\label{Fig: Test Case 1 Comparison}
		\end{figure} \clearpage
		\begin{figure}[hbt!]\centering
			\begin{subfigure}{.85\linewidth}
				\centering
				\includegraphics[width=.85\linewidth]{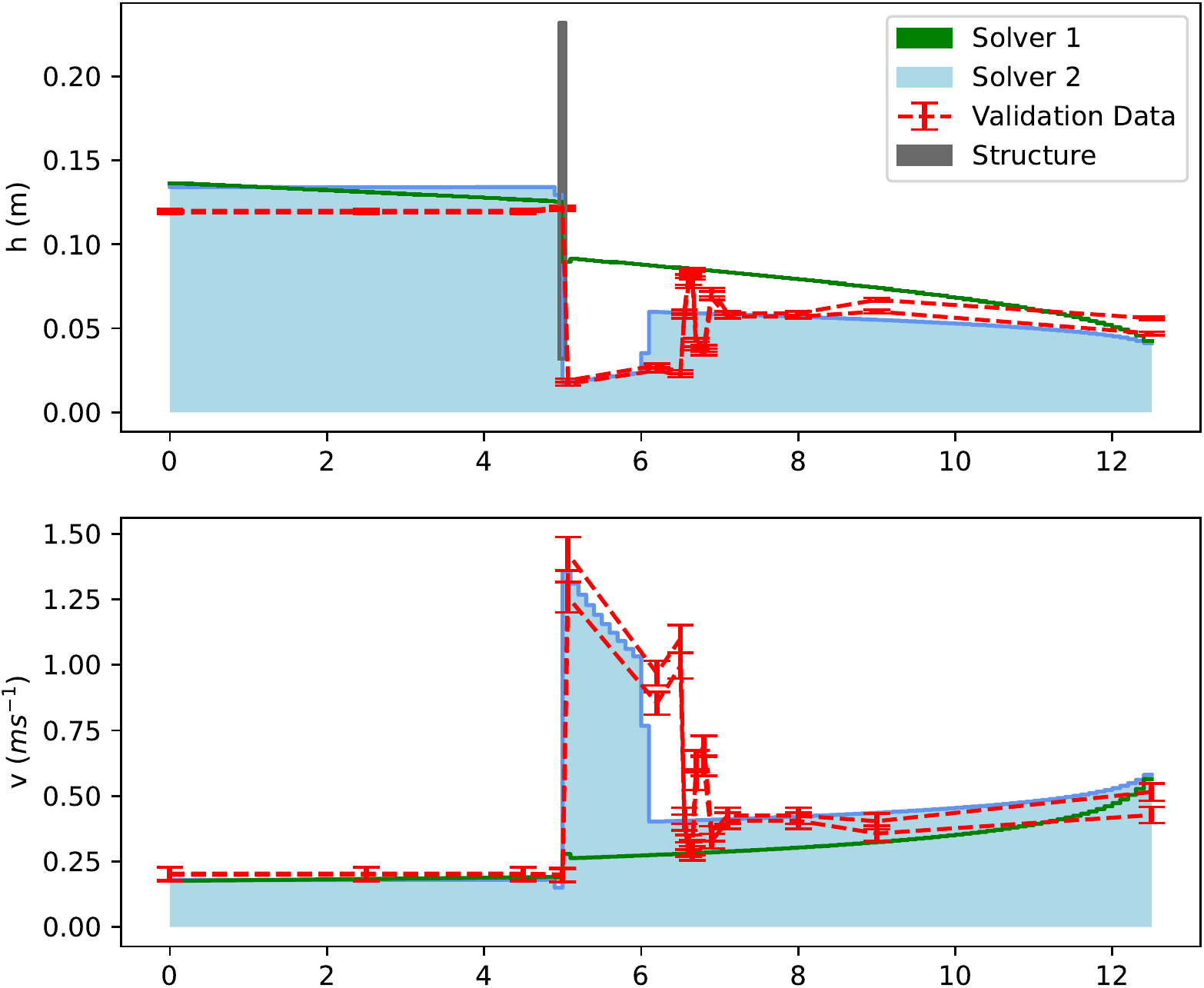}
				\caption{}\label{Fig: Test Case 5 Comparison Plot}
			\end{subfigure}
			\begin{subfigure}{.85\linewidth}
				\centering
				\includegraphics[width=0.85\textwidth]{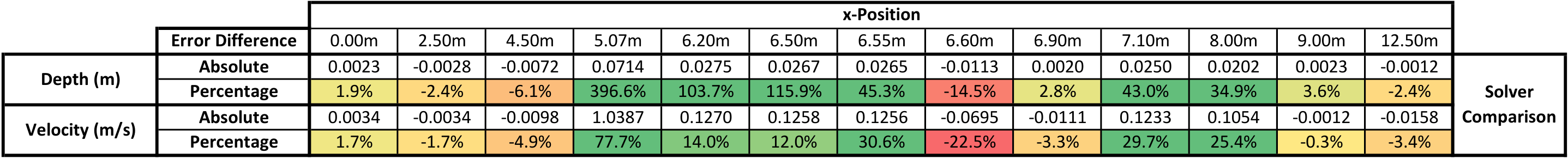}
				\caption{}\label{Fig: Test Case 5 Error Comparison}
			\end{subfigure}
			\caption{Comparison between solver 1 and solver 2 with respect to the experimental results for test case $5$. Details for solver 1 can be found in \cite{RN100}. Details for the numerical setup can be found in Section \ref{section: Numerical Setup}.}
			\label{Fig: Test Case 5 Comparison}
		\end{figure} \clearpage
		\begin{figure}[hbt!]\centering
			\begin{subfigure}{.85\linewidth}
				\centering
				\includegraphics[width=.85\linewidth]{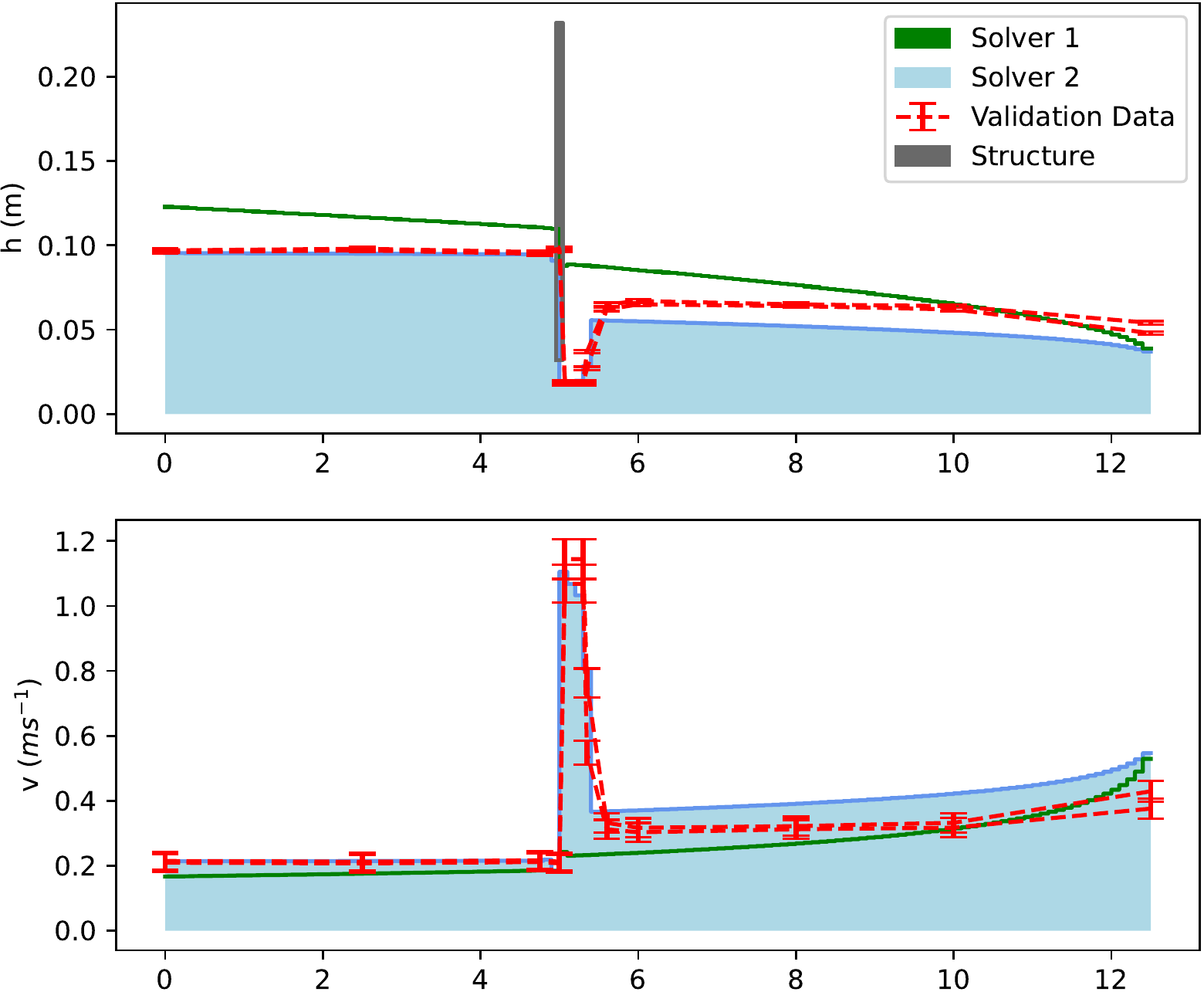}
				\caption{}\label{Fig: Test Case 6 Comparison Plot}
			\end{subfigure}
			\begin{subfigure}{.85\linewidth}
				\centering
				\includegraphics[width=0.85\textwidth]{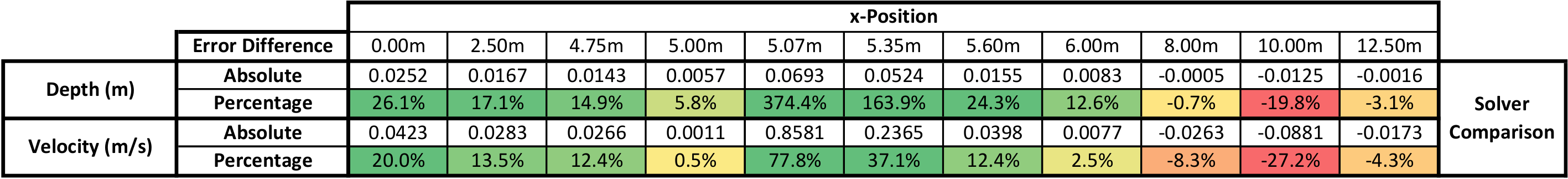}
				\caption{}\label{Fig: Test Case 6 Error Comparison}
			\end{subfigure}
			\caption{Comparison between solver 1 and solver 2 with respect to the experimental results for test case $6$. Details for solver 1 can be found in \cite{RN100}. Details for the numerical setup can be found in Section \ref{section: Numerical Setup}.}
			\label{Fig: Test Case 6 Comparison}
		\end{figure} \clearpage
		\begin{figure}[hbt!]\centering
			\begin{subfigure}{.85\linewidth}
				\centering
				\includegraphics[width=.85\linewidth]{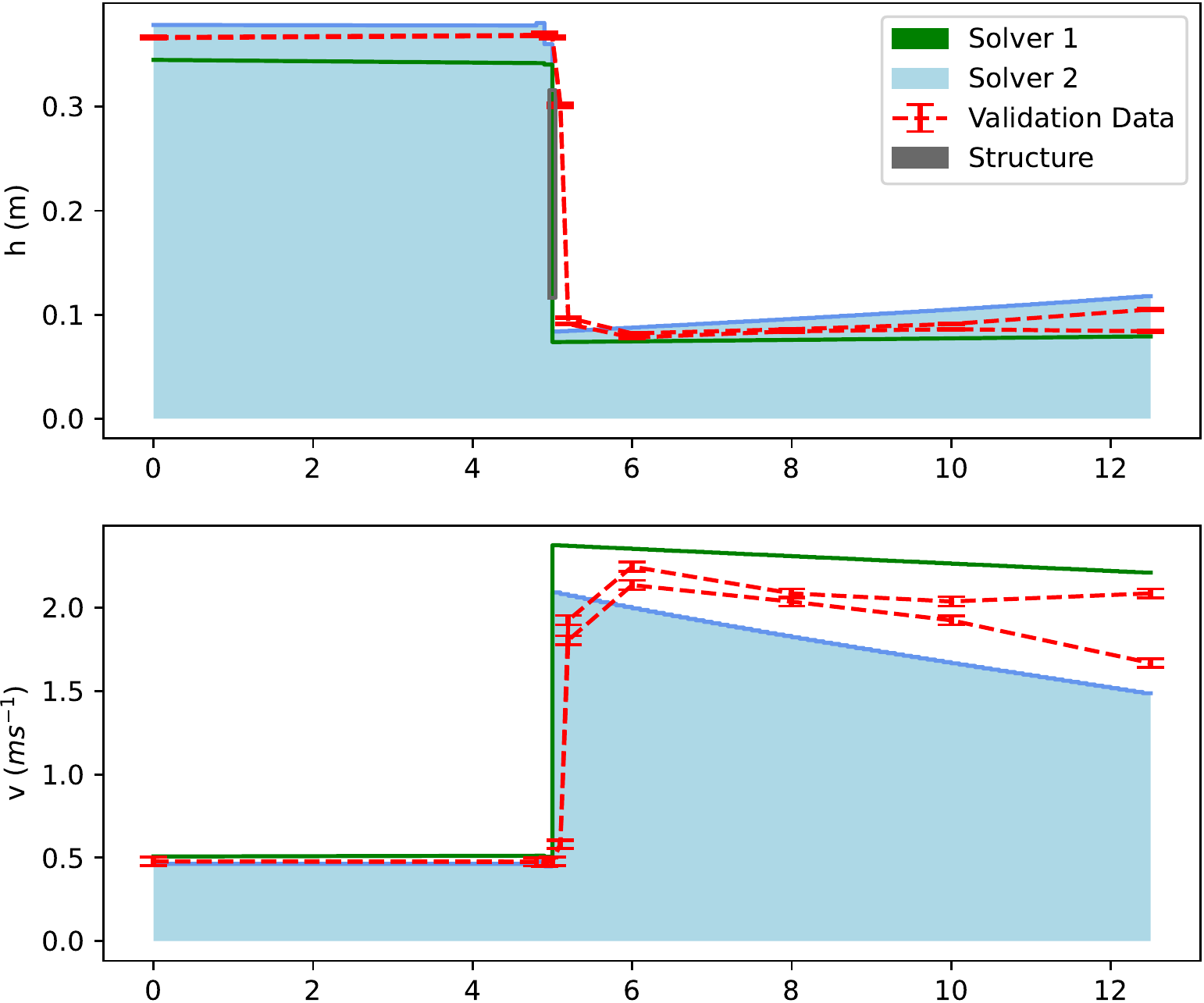}
				\caption{}\label{Fig: Test Case 8 Comparison Plot}
			\end{subfigure}
			\begin{subfigure}{.85\linewidth}
				\centering
				\includegraphics[width=0.85\textwidth]{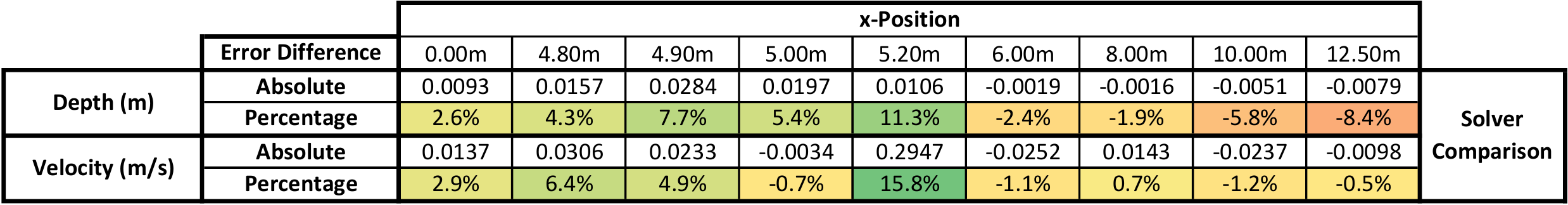}
				\caption{}\label{Fig: Test Case 8 Error Comparison}
			\end{subfigure}
			\caption{Comparison between solver 1 and solver 2 with respect to the experimental results for test case $6$. Details for solver 1 can be found in \cite{RN100}. Details for the numerical setup can be found in Section \ref{section: Numerical Setup}.}
			\label{Fig: Test Case 8 Comparison}
		\end{figure} \clearpage
		\section{Conclusion}
		A new Riemann solver, capable of resolving numerical fluxes across a partially obstructed interface, has been presented. Via the validation process, it has been demonstrated that the solver is able to adequately capture fluid-structure interactions for a range of barrier configurations and flow rates. Furthermore, via the comparison process, it has been demonstrated that the solver represents a significant improvement on the previously published solver \cite{RN100}. It is clear that the new solution procedure addresses the identified weakness of the previous solver by more accurately capturing the vertical variation in the horizontal velocity profile at a structure interface. This results in the more accurate determination of the flow characteristics including the ability to resolve the location and jump height of stationary hydraulic jumps. However, this does come at the cost of increased computational demands and implementation complexity although, due to the local nature of the solution procedure and the proportionally small number of structure cells within a computational domain, the increase in computational expense in unlikely to be prohibitive. As for the previous solver, the biggest barrier to implementation is the scarcity of the required data for structures and the availability of suitable meshing algorithms, which remains the subject of further work. 
		
		The capability of the solver to resolve numerical fluxes across a partially obstructed interface has significant implications for modelling a variety of structures within two-dimensional hydrodynamic models. This has important applications in terms of improving flood inundation modelling capabilities as well as enabling the modelling of infrastructure resilience modelling and the structural health monitoring of hydraulic structures. Moreover, the more accurate determination of the fluid velocity in comparison with the previously presented solver presents new opportunities such as the capacity to model hydraulic jumps and the ability to integrate species equations enabling the modelling of water soluble contaminants in conjunction with flows around obstacles. 
		
		 As for all models, the underlying assumptions must be considered in order to ascertain the limitations of the model and as such the solver should be considered appropriate for modelling structures at a spatial scale whereby approximating the structure as a partial obstruction existing at a cell interface is appropriate. Although, the proposed model does not capture all of the energy losses which occur as a result of the fluid-structure interaction, such effects are insignificant at this spatial scale in comparison with the effect induced by the blockage of the flow by the structure, which is well captured as shown by the validation results. For detailed analyses of individual structures 3D CFD analyses are recommended. 
		 
		 Avenues for further development of the solver are limited without compromising on the compatibility of the solver with standard numerical schemes utilised in contemporary hydrodynamic models. The layer redefinition process, particularly where layers are shifted downwards a significant distance, presents the greatest weakness of the method. However, such cases involve flow which is inherently vertical in nature and it is difficult to consolidate this with the fundamental nature of the shallow water equations which underpin modern hydrodynamic modelling. 
		\section*{Declaration of Competing Interest}
		The authors declare that they have no known competing financial interests or personal relationships that could have appeared to influence the work reported in this paper.
		
		\section*{Acknowledgements}
		This research was funded by the Engineering and Physical Sciences Research Council, United Kingdom grant number EP/T517914/1.
		
		}
		\bibliography{References}
		
		\appendix
		\section{Mesh Convergence Analysis Plots}
		\begin{figure}[hbt!]	
			\begin{center}
				\includegraphics[width=.95\linewidth]{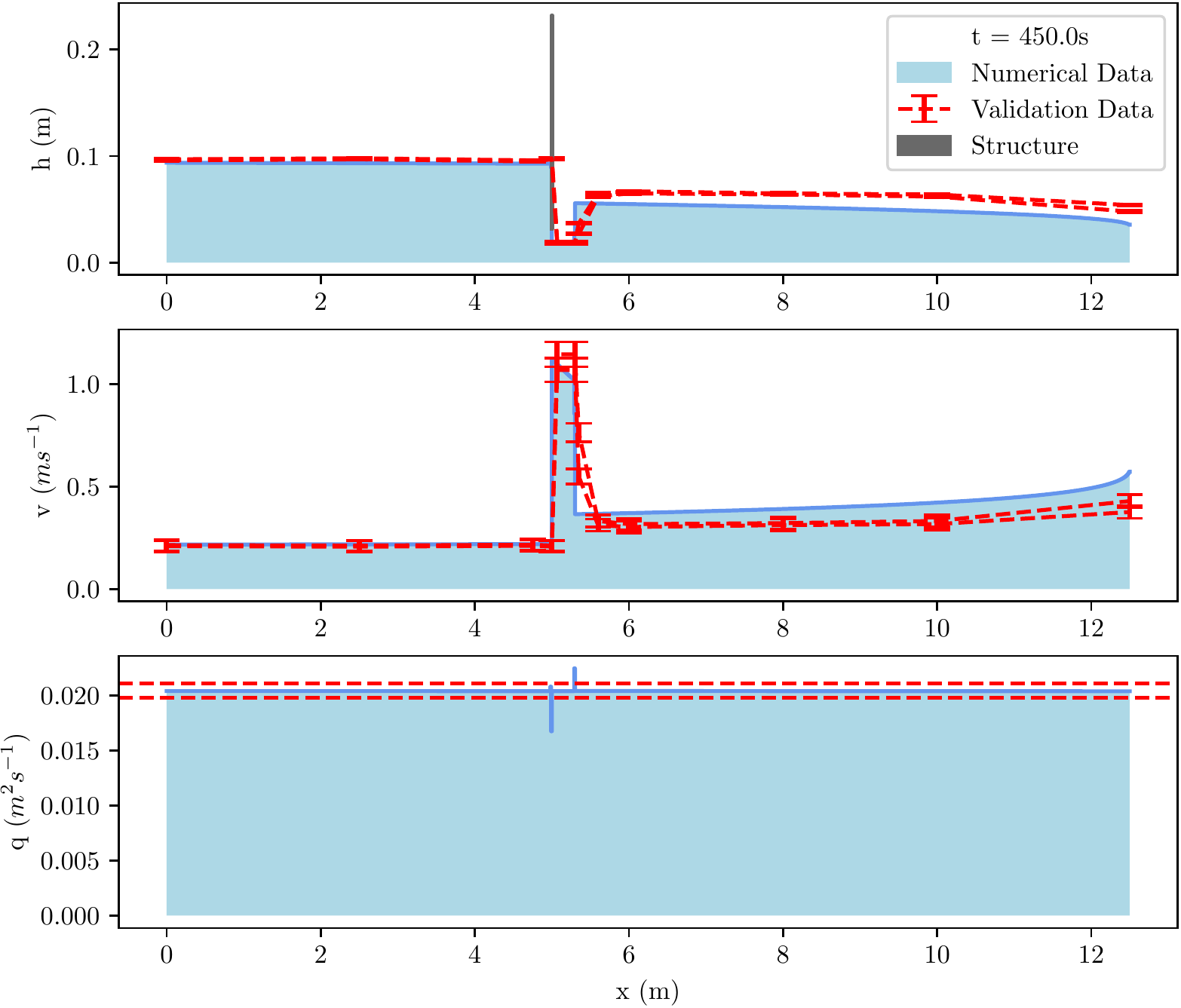}
			\end{center}
			\caption{Numerical results for Test case 6 using a $0.01m$ Grid ($\Delta x = 0.01m$).}
			\label{Fig: MCA 0.01m}
		\end{figure}
		\begin{figure}[hbt!]	
			\begin{center}
				\includegraphics[width=.95\linewidth]{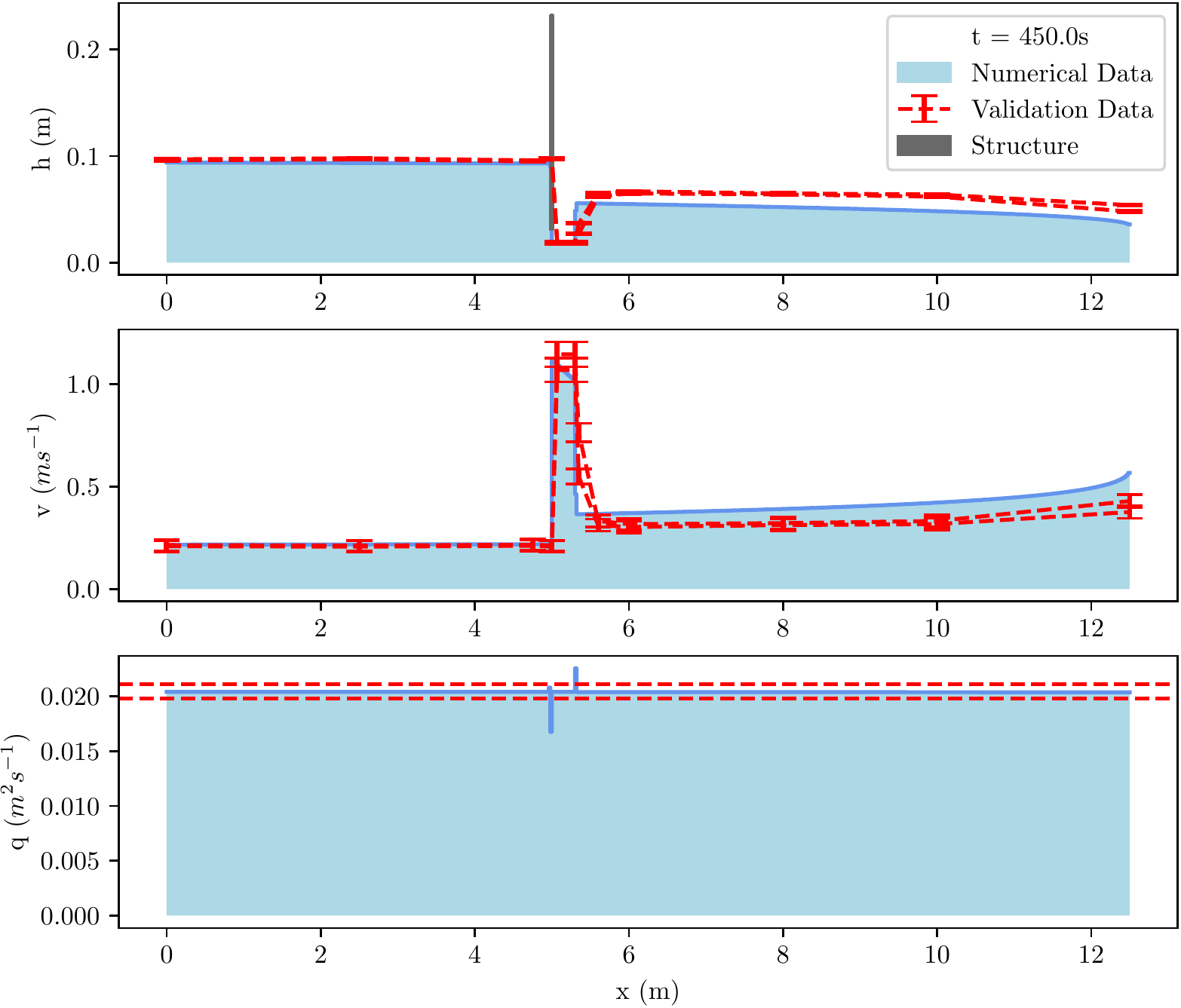}
			\end{center}
			\caption{Numerical results for Test case 6 using a $0.02$m Grid ($\Delta x = 0.02m$).}
			\label{Fig: MCA 0.02m}
		\end{figure}
		\begin{figure}[hbt!]	
			\begin{center}
				\includegraphics[width=.95\linewidth]{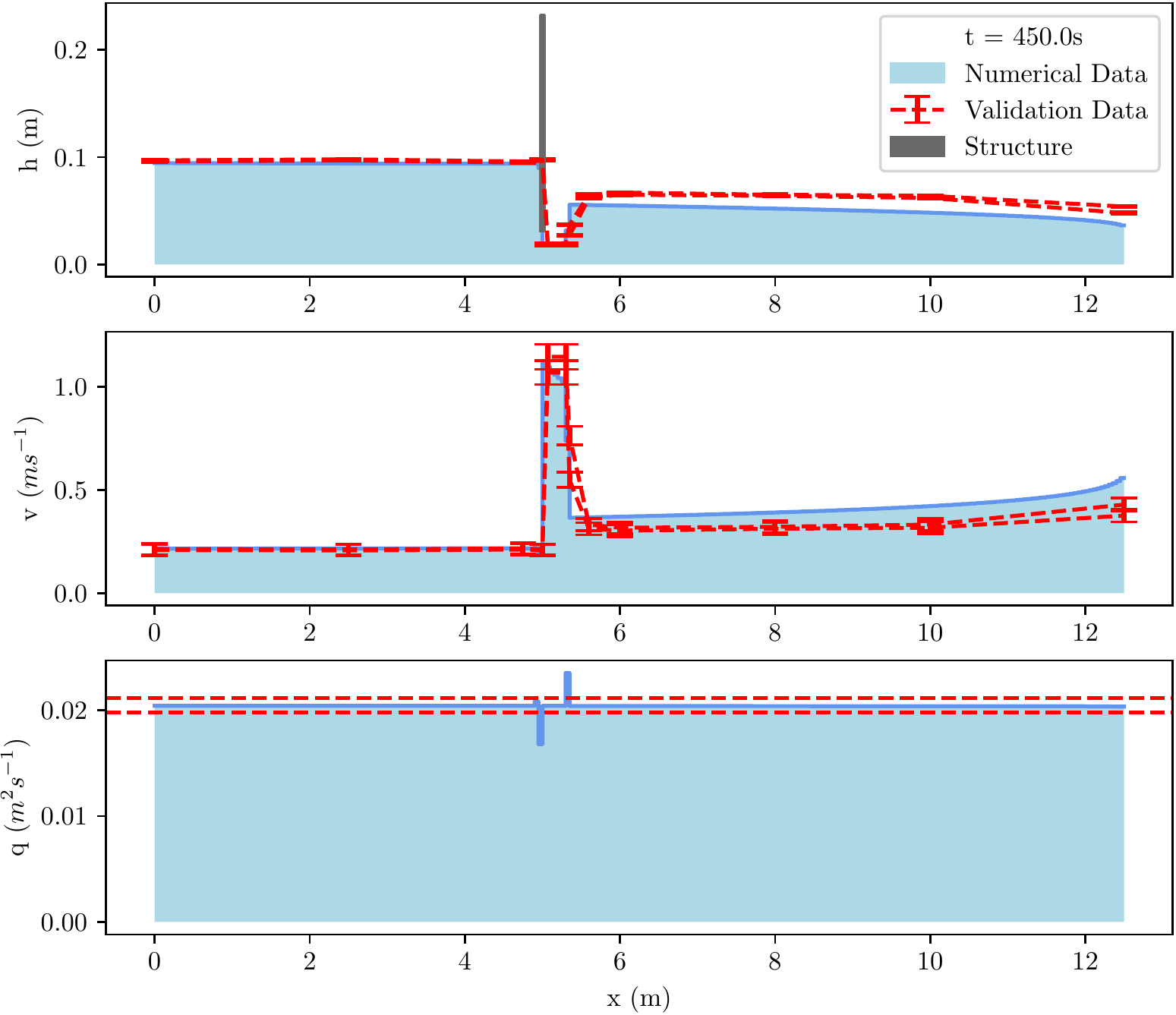}
			\end{center}
			\caption{Numerical results for Test case 6 using a $0.05$m Grid ($\Delta x = 0.05m$).}
			\label{Fig: MCA 0.05m}
		\end{figure}
		\begin{figure}[hbt!]	
			\begin{center}
				\includegraphics[width=.95\linewidth]{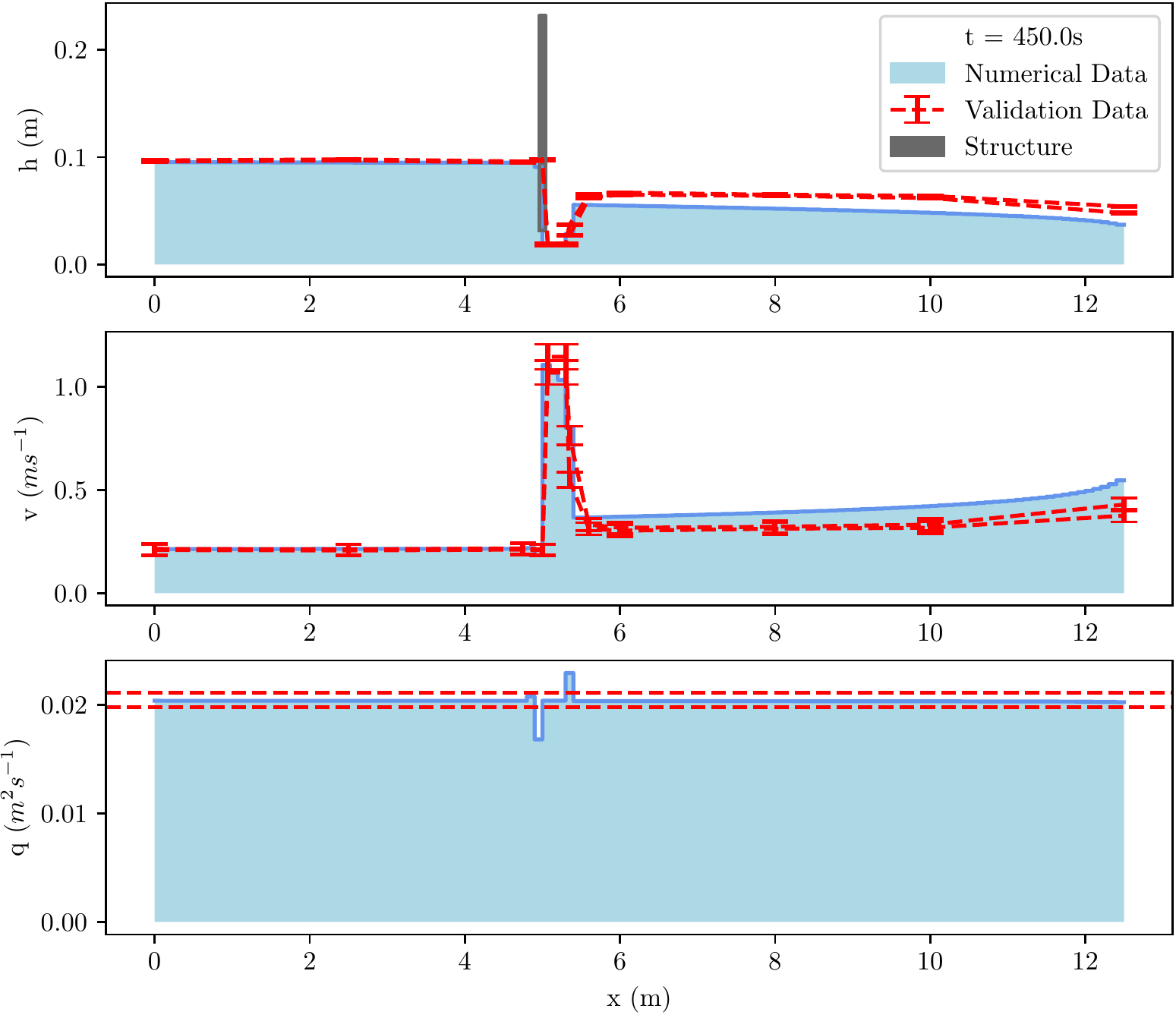}
			\end{center}
			\caption{Numerical results for Test case 6 using a $0.10$m Grid ($\Delta x = 0.10m$).}
			\label{Fig: MCA 0.10m}
		\end{figure}
		\begin{figure}[hbt!]	
			\begin{center}
				\includegraphics[width=.95\linewidth]{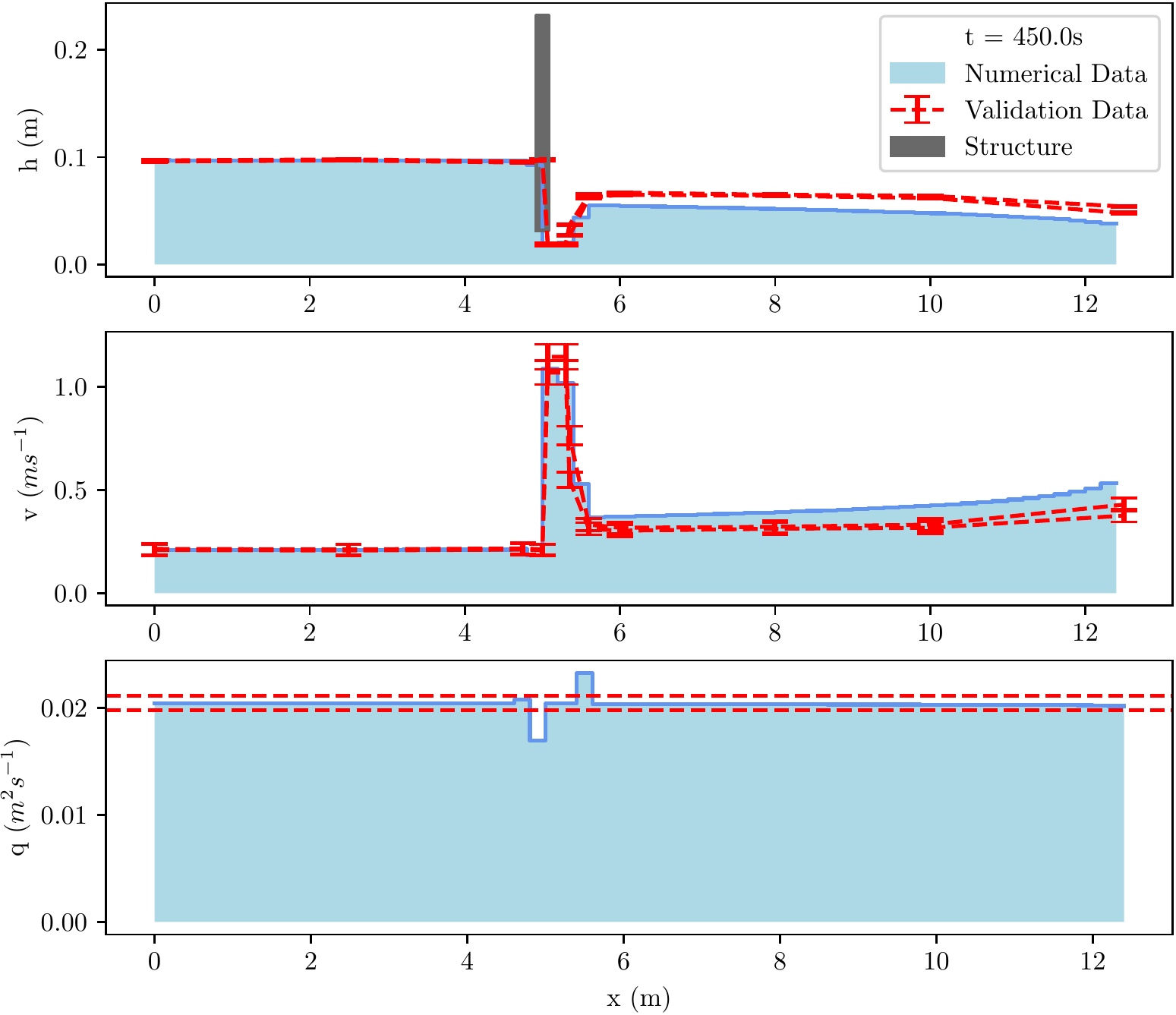}
			\end{center}
			\caption{Numerical results for Test case 6 using a $0.20$m Grid ($\Delta x = 0.20m$).}
			\label{Fig: MCA 0.20m}
		\end{figure}
\end{document}